\documentclass[12pt]{article}

\usepackage[dvips]{graphicx}

\newcommand{\lsim}{\raisebox{-4pt}{$\,\stackrel{\textstyle
                                                         <}{\sim}\,$}}
\newcommand{\gsim}{\raisebox{-4pt}{$\,\stackrel{\textstyle
                                                         >}{\sim}\,$}}
\newcommand{\nn}{\nonumber}
\newcommand{\be}{\begin{equation}}
\newcommand{\ee}{\end{equation}}
\newcommand{\ba}{\begin{eqnarray}}
\newcommand{\ea}{\end{eqnarray}}
\newcommand{\req}[1]{(\ref{#1})}
\def\={\,=\,}
\newcommand{\ci}[1]{\cite{#1}}

\def\mev{~{\rm MeV}}
\def\gev{~{\rm GeV}}

\def\als{\alpha_{\rm s}}
\def\eps{\epsilon}

\def\xbj{x_{\rm Bj}}

\def\xb{\bar{x}}
\newcommand{\tw}{\textwidth}
\def\vk{{\bf k}_{\perp}}

\def\vbs{{\bf b}}
\def\vb0{{\bf b}_0}

\def\xbj{x_{\rm Bj}}

\newcommand{\wf}{wave function}

\def\={\,=\,}

\begin{document} 
\thispagestyle{empty}
\begin{flushright}
WU B 11-06 \\
June, 22 2011\\[20mm]
\end{flushright}

\begin{center}
{\Large\bf Transversity in hard exclusive electroproduction of pseudoscalar
  mesons} \\
\vskip 10mm

S.V.\ Goloskokov
\footnote{Email:  goloskkv@theor.jinr.ru}
\\[1em]
{\small {\it Bogoliubov Laboratory of Theoretical Physics, Joint Institute
for Nuclear Research,\\ Dubna 141980, Moscow region, Russia}}\\
\vskip 5mm

P.\ Kroll \footnote{Email:  kroll@physik.uni-wuppertal.de}
\\[1em]
{\small {\it Fachbereich Physik, Universit\"at Wuppertal, D-42097 Wuppertal,
Germany}}\\
and\\
{\small {\it Institut f\"ur Theoretische Physik, Universit\"at
    Regensburg, \\D-93040 Regensburg, Germany}}\\

\end{center}
\vskip 5mm 
\begin{abstract}
Estimates for electroproduction of pseudoscalar mesons at small values of
skewness are presented. Cross sections and asymmetries for these processes
are calculated within the handbag approach which is based on factorization 
in hard parton subprocesses and soft generalized parton distributions (GPDs). 
The latter are constructed from double distributions. Transversity GPDs are 
taken into account; they are accompanied by twist-3 meson wave functions. 
For most pseudoscalar-meson channels a combination of $\widetilde{H}_T$ and 
$E_T$ plays a particularly prominent role. This combination of GPDs which we 
constrain by moments obtained from lattice QCD, leads with the exception of 
the $\pi^+$ and $\eta^\prime$ channels, to large transverse cross sections. 
\end{abstract}

\section{Introduction}
\label{sec:intro}

In a recent paper \ci{GK5} we have investigated hard exclusive 
electroproduction of positively charged pions within the framework of the
handbag approach in the kinematical range of small skewness, $\xi$, and small
invariant momentum transfer, $-t$ but large photon virtualities, $Q^2$. As in
our previous studies of vector-meson electroproduction \ci{GK2,GK3} the
partonic subprocess is calculated within the modified perturbative approach
\ci{botts89} in which quark transverse degrees of freedom as well as Sudakov
suppressions are taken into account. In other words the transverse size of the 
produced meson is not ignored as in the collinear (leading-twist) approach. On 
the other hand, the partons entering the subprocess are viewed as being
emitted and reabsorbed by the nucleon collinearly to the nucleon momenta. The 
GPDs which embody the soft physics, are 
constructed with the help of double distributions. In \ci{GK5} the sketched 
approach has been used to analyze the HERMES data on $\pi^+$ electroproduction 
\ci{hermes07,Hristova,hermes02}. Besides the use of the modified perturbative
approach this analysis differs in the following aspects from previous studies
based on the collinear approximation (e.g.\ \ci{man98} - \ci{schafer05}):
\begin{itemize}
\item The full electromagnetic form factor of the pion, $F_\pi(Q^2)$, is 
  taken into account (see also Ref.\ \ci{bechler}) as it is measured by the 
  $F_\pi$ collaboration \ci{blok} at Jefferson Lab in just the same process
\item Besides the pion pole there is an extra contribution to the
  GPD $\widetilde{E}$. 
\item There are substantial contributions from transversely polarized
  photons. This is particularly obvious from the $\sin{\phi_s}$-moment, 
  $A_{UT}^{\sin{\phi_s}}$, of the $\pi^+$ cross section measured with a
  transversely polarized target \ci{Hristova}. As it is argued in \ci{GK5} 
  the contributions from transversely polarized photons can be calculated 
  within the handbag approach as a twist-3 effect consisting of a twist-3 
  pion \wf{} and the leading-twist transversity GPD $H_T$. 
\end{itemize}

Here, in this work we are going to extend the analysis performed in \ci{GK5}
to other pseudoscalar meson channels, namely $\pi^0$, $K$ and $\eta
(\eta^\prime)$ production. For these processes there are no small-skewness
data available as yet but they may be measured at the upgraded Jefferson Lab
facility or by the Compass experiment. In so far we believe that estimates
for various pseudoscalar meson channels are of interest and timely. While for
$\pi^+$ production only the isovector combinations of the GPDs, 
$F^{(3)}=F^u-F^d$ ($F=\widetilde{H}, \widetilde{E}, H_T$), contribute,
different flavor combinations are relevant for the other channels. This 
necessitates a careful reexamination of the parameterizations of the GPDs with 
particular regard to those of the individual flavors. In some cases, as for 
instance for $\widetilde{E}$, we have to revise slightly the parameterizations
proposed in \ci{GK5}. In a recent lattice-QCD study \ci{goeckeler} large 
moments of the transversity GPD combination
\be
\bar{E}_T \= 2 \widetilde{H}_T + E_T
\label{eq:combination}
\ee
has been found. With regard to this result we will also examine the role of 
this GPD which has not been taken into account in \ci{GK5}, in order to find
out whether or not it provides substantial effects in observables for
meson electroproduction.

The plan of the paper is the following: In the next section we will sketch 
the handbag approach to meson electroproduction including twist-3 effects. In 
Sect.\ \ref{sec:GPD} we are going to present the parameterizations of the GPDs 
and compare them with recent results from lattice QCD in detail. Predictions
for electroproduction of pseudoscalar mesons will be presented in  the
subsequent sections, for pions in Sec.\ \ref{sec:pion}, for $\eta$ and 
$\eta^\prime$ in Sect.\ \ref{sec:eta} and for kaons in Sect.\ \ref{sec:kaon}. 
In Sect.\ \ref{sec:supp} the asymmetries obtained with either a longitudinally 
polarized beam or target will be discussed. The paper is closed with a summary 
(Sect.\ \ref{sec:summary}). 
      
\section{An outline of the handbag approach}
\label{sec:handbag}
For details of the approach we are going to use it is referred to our previous
papers, e.g.\ \ci{GK5,GK2,GK3}. Here, we only sketch the basic facts. It is to
be stressed that we consider the kinematical region of small $\xi$ and
small $-t$ but large $Q^2$ and large photon-proton c.m.s.\ energy, $W$. 
Terms of order $(\sqrt{-t}/Q)^n$ ($n\geq 1$) are neglected throughout.

The helicity amplitudes for electroproduction of pseudoscalar mesons,
$\gamma^* p\to PB$, through longitudinally polarized photons read
\ba
{\cal M}^{P}_{0+,0+} &=& \sqrt{1-\xi^2}\, \frac{e_0}{Q}
                             \,\Big[\langle \widetilde{H}^{P}\rangle
-\frac{\xi^2}{1-\xi^2}\langle \widetilde{E}^{P}_{\rm n.p.}\rangle 
  - \frac{\xi(m+M)Q^2}{1-\xi^2}\frac{\rho_P}{t-m_P^2}\Big]\,,\nn\\
{\cal M}^{P}_{0-,0+} &=& \frac{e_0}{Q}\,\frac{\sqrt{-t^\prime}}{m + M}\,\Big[ \xi 
\langle \widetilde{E}^{P}_{\rm n.p.}\rangle + (m+M)Q^2\frac{\rho_P}{t-m_P^2}\Big]\,.
\label{eq:L-amplitudes}
\ea 
Helicities are labeled by their signs or by zero. The usual abbreviation 
$t^\prime=t-t_0$ is employed where \ci{schafer05} 
\be
t_0=-2\, \frac{(m^2+M^2)\xi^2+(M^2-m^2)\xi}{1-\xi^2} 
\ee
is the minimal value of $-t$ corresponding to forward scattering. The mass of
the nucleon (meson, final state baryon) is denoted by $m (m_P, M)$, $e_0$ is
the positron charge and the skewness is related to Bjorken-$x$ by 
\be
\xi\=\frac{\xbj}{2-\xbj}\big[ 1 + m_P^2/Q^2\big]\,.
\ee

\begin{figure}[t]
\begin{center}
\includegraphics[width=0.60\tw,bb=103 515 568 656,clip=true]{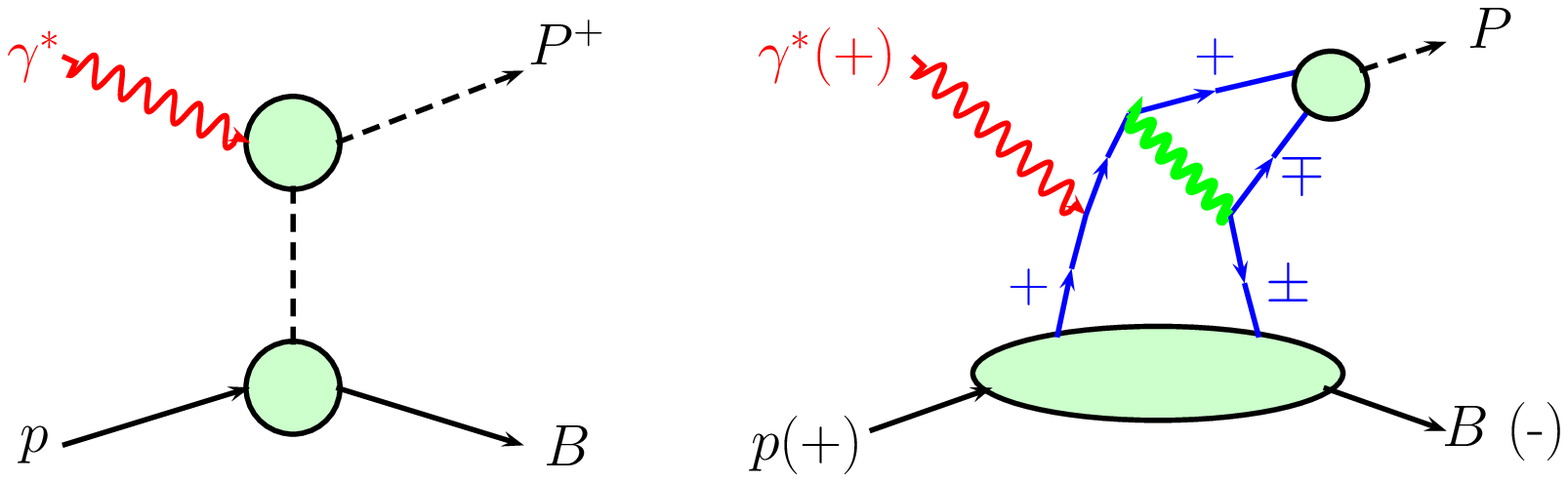}
\caption{\label{fig:graphs} The pole contribution to the process $\gamma^*p\to
  P^+B$(left) and the handbag graph (right) for
  electroproduction of pseudoscalar mesons. Helicities are specified for the
  amplitude ${\cal M}_{0-,++}$.}
\end{center}
\end{figure}

The pole contribution, see Fig.\ \ref{fig:graphs}, occuring in $\pi^+$ and
$K^+$ production  has the residue 
\be
\rho_P\= g_{PpB}F_{PpB}(t) F_P(Q^2)\,,
\label{eq:residue}
\ee
where $g_{PpB}F_{PpB}(t)$ is the coupling of the meson to the proton-baryon
vertex and $F_P$ represents the electromagnetic form factor of the meson
for which we use the experimental values. For $\pi^0$ and $\eta$ production 
$\rho_P$ is zero. Last not least the item $\langle F^P\rangle$ in 
\req{eq:L-amplitudes} denotes a convolution of the GPD $F$ with an appropriate 
subprocess amplitude to be calculated from a set of Feynman graphs of which 
a typical leading-order example is shown in Fig.\ \ref{fig:graphs},
\be
\langle F^{P} \rangle \= \sum_\lambda\int_{-1}^1 d\xb\, 
   {\cal H}^P_{0\lambda,0\lambda}(\xb,\xi,Q^2,t=0) F^{P}(\xb,\xi,t)\,.
\label{eq:convolution}
\ee 
The label $\lambda$ refers to the unobserved helicities of the partons
participating in the subprocess.  
Within the modified perturbative approach the subprocess amplitude reads 
\ba
{\cal H}^P_{0\lambda,0\lambda} &=& \int d\tau d^2b\, 
         \hat{\Psi}_{P}(\tau,-\vbs,\mu_F)\, 
      \hat{\cal F}^{P}_{0\lambda,0\lambda}(\xb,\xi,\tau,Q^2,\vbs,\mu_R)\, \nn\\ 
      && \times   \als(\mu_R)\,{\rm exp}{[-S(\tau,\vbs,Q^2,\mu_F,\mu_R)]}
\label{mod-amp}
\ea
in the impact parameter (${\bf b}$) space. For the Sudakov factor $S$, the 
choice of the renormalization ($\mu_R$) and factorization ($\mu_F$) scales as
well as the hard scattering kernels ${\cal F}$ or their respective Fourier 
transforms $\hat{\cal F}$, we refer to Ref.\ \ci{GK3}. The last item in 
\req{mod-amp} to be explained is $\hat{\Psi}_{P}(\tau,-\vbs)$ which represents
the Fourier transform of the momentum-space light-cone \wf{} for the
meson ($\tau$ is the momentum fraction of the quark that enters the meson, 
defined with respect to the meson momentum, the antiquark carries the fraction
$1-\tau$). 

As forced by the small $-t^\prime$ behavior of the HERMES data \ci{Hristova}
on $A_{UT}^{\sin{\phi_s}}$ for $\pi^+$ electroproduction one also needs the
amplitude ${\cal M}_{0-,++}$. As is obvious from the parton helicities
specified in the Feynman graph depicted in Fig.\ \ref{fig:graphs}, the usual
GPDs $\widetilde{H}$ and $\widetilde{E}$ which parameterize the nucleon matrix
element for a situation where the emitted and reabsorbed partons have the same
helicity, do not provide a contribution to ${\cal M}_{0-,++}$ with the
required behavior ${\cal M}_{0-,++}\to {\rm const.}$ for $t^\prime\to 0$. 
Angular momentum conservation forces this contribution to vanish in the
forward direction. What is required, as has been discussed in detail in 
\ci{GK5}, is a contribution from the helicity-flip or transversity GPD $H_T$ 
\ci{hoodbhoy,diehl01} in combination with a twist-3 meson \wf{}   
\be
{\cal M}^P_{0-,++}\= e_0\sqrt{1-\xi^2}\,\int_{-1}^1\,d\xb\, {\cal H}^{P\,
  {\rm\scriptscriptstyle twist-3}}_{0-,++} \, H_T^P\,.
\label{eq:ht-ampl}
\ee
This amplitude is parametrically suppressed by $\mu_P/Q$ with respect to the
asymptotically leading amplitudes for longitudinally polarized photons. The
parameter $\mu_P$ is large since it is enhanced by the chiral condensate
\be
\mu_P \= \frac{m_P^2}{m_1+m_2}\,,
\label{eq:chiral}
\ee
by means of the divergency of the axial-vector current. The $m_i$ are the
current-quark masses of the meson's valence quarks. The parameter $\mu_P$ is
scale dependent and evolves with the anomalous dimensions $4/\beta_0=12/25$ 
for four flavors. For the $\eta$ and $\eta^\prime$ the situation is a bit more
complicated. Decomposing these mesons into flavor-octet and single states, 
one has for the octet case 
\be
\mu_{\eta_8}\=3\,\frac{m_{\eta_8}^2}{m_u+m_d+4m_s}\,.
\label{eq:chiral-eta}
\ee
The flavor-singlet $\eta$ is not related to the chiral condensate. In this
case $\mu_{\eta_1}$ is just the $\eta_1$ mass which we approximate by that
of the $\eta^\prime$. For pions, kaons and $\eta_8$ we take a value of 
$2\,\gev$ for the parameter $\mu_p$ at a scale of $2\,\gev$. 

The twist-3 subprocess amplitude is given explicitly in \ci{GK5} for $\pi^+$
electroproduction. Its generalization to other pseudoscalar mesons is
straightforward.  
 
As we mentioned in the introduction we will also examine the role of the GPD
\req{eq:combination} which contributes to the amplitude
\be
{\cal M}^P_{0+,\mu +}\= -\frac{e_0}{2}\frac{\sqrt{-t^\prime}}{m + M}\,
     \int_{-1}^1\,d\xb\, 
     {\cal H}^{P\,{\rm\scriptscriptstyle twist-3}}_{0-,++} \, \bar{E}_T^P\,,
\label{eq:ebar-ampl}
\ee  
where $\mu=\pm 1$ indicates a transverse photon helicity. For our kinematical
range of small $-t^\prime$ and small skewness contributions from the other
transversity GPDs to the amplitudes \req{eq:ht-ampl} and \req{eq:ebar-ampl}
can be neglected. Note that $\widetilde{E}_T$ is an odd function of $\xi$ as a
consequence of time-reversal invariance \ci{diehl01}. The double-flip
amplitude ${\cal M}_{0-,-+}$ is also suppressed in the kinematical range of
interest and neglected.

The pion-pole contributions to the amplitudes for transversely polarized
photons are also taken into account. Explicit expressions for these
contributions can be found in \ci{GK5}.
\section{The parameterization of the GPDs}
\label{sec:GPD}
According to the discussion presented in Sect.\ \ref{sec:handbag} we need to
model the GPDs $\widetilde{H}$, $\widetilde{E}$, $H_T$ and $\bar{E}_T$. It is
to be emphasized that our parameterizations of the various GPDs  are optimized
for small skewness and small $-t$. We assume flavor-symmetric sea GPDs
throughout. Hence, we have to model only the valence-quarks GPDs since
the sea contributions cancel for $\pi^+$ and $K^+$ production on this 
assumption. For the cases of $\pi^0$ and $\eta$ the sea does anyway not
contribute. At present there is no experimental information from hard
exclusive processes available which would allow for an examination of 
flavor-symmetry breaking effects in the sea. In recent studies of the 
polarized parton distributions \ci{leader,vogelsang} however flavor-symmetry 
breaking effects in the sea have been found. Thus, at least for $\widetilde{H}$ 
an estimate of the sea contribution on the basis of the double distribution 
ansatz, is possible. We found its sea contribution to the $\pi^+$ production 
cross section to amount to about $1\%$. It therefore can safely be neglected.

The valence-quarks GPDs are constructed with the help of their double
distribution representation. For the latter the familiar ansatz \ci{mus99}
\be
f_i^a (\rho, \eta,t) \= \exp{[(b_i-\alpha^\prime_i\ln{\rho})t]}\, 
        F_i^a(\rho,\xi=t=0)\,
      \frac{3}{4}\frac{[(1-\rho)^2-\eta^2]}{(1-\rho)^3}\,\Theta{(\rho)} 
\label{eq:DD}
\ee
is made which consists of the forward limit, $F_i$, of the relevant GPD, a
weight function and an exponential in $t$ with a profile function
parameterized in a Regge-like fashion, i.e.\  with a slope of a Regge
trajectory, $\alpha_i^\prime$, and a slope, $b_i$, of the residue function. The
$t=0$ part of the Regge term is absorbed in the forward limit of the GPD
which is given either by the polarized  ($\Delta q^a(x)$) or the transversity
($\delta^a(x)$) parton distribution functions (PDFs) for quarks of flavor $a$ 
for the GPDs $\widetilde{H}^a$ or $H_T^a$. For the remaining two GPDs, 
$\widetilde{E}^a$ and $\bar{E}_T^a$, the forward limit is parameterized in a 
fashion analogously to the PDFs
\be
F_i^a(\rho,\xi=t=0)\=N_i^a \rho^{-\alpha_i(0)}\,(1-\rho)^{\beta_i^a}
\label{eq:pdf}
\ee
with the parameters to be adjusted appropriately.  

The full GPDs are obtained from the double distributions by the integral
\ci{mul94,rad98}
\be
F^a_i(\xb,\xi,t)=\int_{-1}^1\,d\rho\,\int_{-1+|\rho|}^{1-|\rho|}\, 
                    d\eta\,\delta(\rho+\xi\eta-\xb)\,f^a_i(\rho,\eta,t)\,.
\label{eq:integral}
\ee

Evolution is neglected throughout since we are only interested in scales
in the proximity of $2\,\gev$ which is the scale at which we quote 
the parameterizations of the GPDs. Below we will occasionally refer to moments 
of GPDs, in particular to moments at zero skewness. They are defined by
\be
F_{in0}(t)\=\int_{-1}^1 d\xb\, \xb^{n-1}\,F_i(\xb,\xi=0,t)\,.
\label{eq:moments}
\ee

Before we turn to the discussion of the individual GPDs we want to comment on
the Regge trajectories, $\alpha_i(t)$, used by us. They are to be understood
as effective trajectories describing unspecified superpositions of poles and 
cuts. They are not related to the hadronic spectrum as for instance the
prominent $\rho$ and $\omega$ Regge trajectories which contribute to the GPDs 
$H$ and $E$. The appearance of cuts in soft reactions dominated by
unnatural-parity exchanges is a well-known fact. For definiteness let us 
consider the closely related reaction photoproduction of mesons ($M$). For 
single-particle exchanges the corresponding helicity amplitudes satisfy the 
symmetry relation~\footnote{
In \ci{GK5,GK3} it is shown that this relation also holds for the handbag
amplitude at the twist-2 level (with or without power corrections like $\vk$
effects and to any order of perturbative QCD). The contributions from the GPDs 
$H$ and $E$ ($\widetilde{H}$ and $\widetilde{E}$) corresponds to (un)natural 
parity exchange. Contributions from $\bar{E}_T$ behaves as natural parity 
exchanges while those from $H_T$ do not possess the property \req{eq:symmetry}.}
\be
{\cal M}^M_{-\mu^\prime \nu^\prime,-\mu\nu}\=\kappa \eta_M
(-1)^{\mu-\mu^\prime}\,{\cal M}^M_{\mu^\prime \nu^\prime,\mu\nu}\,,
\label{eq:symmetry}
\ee
where, for vector (pseudoscalar) mesons, $\eta_M=+(-)1$ and, for a (un)natural 
parity exchange, $\kappa=+(-)1$. Eq.\ \req{eq:symmetry} relates for instance the
amplitudes ${\cal M}_{0-,-+}$ and ${\cal M}_{0-,++}$. The first one is a 
double-flip amplitude which, by angular momentum conservation, is forced to 
vanish $\propto t^\prime$ for forward scattering. Therefore, ${\cal M}_{0-,++}$ 
also vanishes $\propto t^\prime$ for dynamical reasons although it is a 
non-flip amplitude. Hence, for $\pi^+$ photoproduction, all single-particle 
exchanges vanish for forward scattering and would produce a forward dip in the 
differential cross section. This is in sharp contrast to experiment
\ci{boyarski} - the $\pi^+$ photoproduction cross section exhibits a pronounced 
forward spike.

\subsection{$\widetilde{H}$}
The forward limit of $\widetilde{H}$, the polarized parton
distribution~\footnote{
A suitable parameterization of these PDFs can be found in \ci{GK3}. As
compared to that work the parameterization of $\Delta u_v$ is changed somewhat.} 
is taken from \ci{BB}. The Regge parameters for $\widetilde{H}$ are quoted in
Tab.\ \ref{tab:1}. In \ci{DFJK4} a profile function for $\widetilde{H}$ has
been proposed that is more complicated than 
$b_{\tilde{H}}-\alpha_{\tilde{H}^\prime} \ln{\rho}$ in \req{eq:DD}. It reads
\be
-\alpha_{\tilde{H}}^\prime (1-\rho)^3\ln{\rho} + b_{\tilde{H}}^a(1-\rho)^3 
          + A_{\tilde{H}}^a \rho (1-\rho)^2
\label{eq:dfjk4}
\ee
In \ci{DFJK4} the parameters appearing in \req{eq:dfjk4}, have been determined
from the data on the axial-vector form factor \ci{kitagaki} through the sum
rule for $\widetilde{H}$:
\ba
\alpha_{\tilde{H}}^\prime&=&0.9\,\gev^{-2}\,, \qquad
b_{\tilde{H}}^u\=b_{\tilde{H}}^d\=0.59\,\gev^{-2}\,\nn \\
A_{\tilde{H}}^u &=&1.22\,\gev^{-2}\, \qquad A_{\tilde{H}}^d =2.59\,\gev^{-2}\,.
\ea
This profile function is also valid at large $-t$. As has been noted in
\ci{DFJK4} there is a strong correlation between $\rho$ and $t$ in the
zero-skewness GPDs: at large (small) $-t$ the large (small) $\rho$-region
dominates. With regard to this fact one can view the profile function in
\req{eq:DD} as the small $-t$ approximation of the more complicated profile 
function \req{eq:dfjk4}. Indeed both the profile functions for $\widetilde{H}$ 
lead to practically the same results for the axial-vector form factor and for
pion electroproduction at small $-t$. In Fig.\ \ref{fig:FF} the first moment 
of the isovector combination of $\widetilde{H}$ which represents the axial
form factor of the nucleon
\be
F_A(t) \= \int_{-1}^1 d\xb\, \widetilde{H}^{(3)}(\xb,\xi,t)\,,
\ee
is shown and compared to the experimental data \ci{kitagaki} and to results
obtained from a recent lattice-QCD simulation \ci{lattice07}. Agreement
between the moments evaluated from $\widetilde{H}$ and experiment is
to be observed while the lattice result exhibits a flatter $t$ dependence
although it corresponds with the others at small $-t$. This flat $t$
dependence is probably related to the fact that the lattice-QCD results are 
evaluated from heavy quarks (corresponding to a pion mass of $352\,\mev$), 
the extrapolation to the chiral limit is not yet possible in general. Similar 
observations can also be made for higher moments and for the moments of the 
isoscalar GPD combination. In passing we remark that an analogous comparison 
for the Dirac and Pauli form factors \ci{DFJK4} (or higher moments of $H$ and 
$E$) reveals similar results. It is also to be stressed that, with regard to 
the heavy pion used in present lattice calculations, utmost caution is 
advisable in any comparison between lattice and phenomenological results.
\begin{table*}[t]
\renewcommand{\arraystretch}{1.4} 
\begin{center}
\begin{tabular}{| c || c | c | c || c | c |}
\hline   
GPD & $\alpha(0)$ & $\alpha^\prime [\gev^{-2}]$ & $b [\gev^{-2}]$ & $N^u$ &
$N^d$ \\[0.2em]  
\hline
$\widetilde{H}$ & 0.48 & 0.90  &  0.59  & -  & - \\[0.2em]
$\widetilde{E}_{\rm n.p.}$ & 0.48 & 0.45 & 0.9 & 14.0 & 4.0 \\[0.2em]
$H_T$ & -0.02 & 0.45 & 0 & 0.78 & -1.01 \\[0.2em]
$\bar{E}_T$& 0.3 & 0.45 & 0.5 & 6.83 & 5.05 \\[0.2em]
\hline
\end{tabular}
\end{center}
\caption{Regge parameters and normalizations of the GPDs, quoted at a scale
  of $2\,\gev$.}
\label{tab:1}
\renewcommand{\arraystretch}{1.0}   
\end{table*} 

\subsection{$\widetilde{E}$}
As already mentioned this GPD consists of two parts, a contribution from the
meson-pole and a non-pole term; the latter has been ignored in most studies
(cf.\ for instance \ci{man98,frankfurt99,goeke00,schafer05}). The pole term
can be written as \ci{goeke00} 
\be
\widetilde{E}^{\rm pole}_{p\to B} \= \Theta(\mid\xb\mid\leq\xi)\,
   \frac{F_{PB}^{\rm pole}(t)}{2\xi}\,\Phi_P\left(\frac{\xb+\xi}{2\xi}\right)
\label{eq:Etilde-pole}
\ee
where $F_{PB}^{\rm pole}$ is the pole contribution to the pseudoscalar form
factor for the $p\to B$ transition~\footnote{
The meson-exchange pole also contributes to the transversity GPDs $H_T$, $E_T$
and $\widetilde{E}_T$. The pole contribution to the amplitudes for 
transversely polarized photons are non-zero.}. 
The non-pole contribution, $\widetilde{E}_{\rm n.p.}$, to $\widetilde{E}$ is
parameterized as in \req{eq:DD} - \req{eq:integral}. The parameters are quoted
in Tab.\ \ref{tab:1}. The power $\beta^u_{\tilde{E}}=\beta^d_{\tilde{E}}$ is
taken as 5.

\begin{figure}[t]
\begin{center}
\includegraphics[width=0.45\tw,bb=158 370 589 720,clip=true]{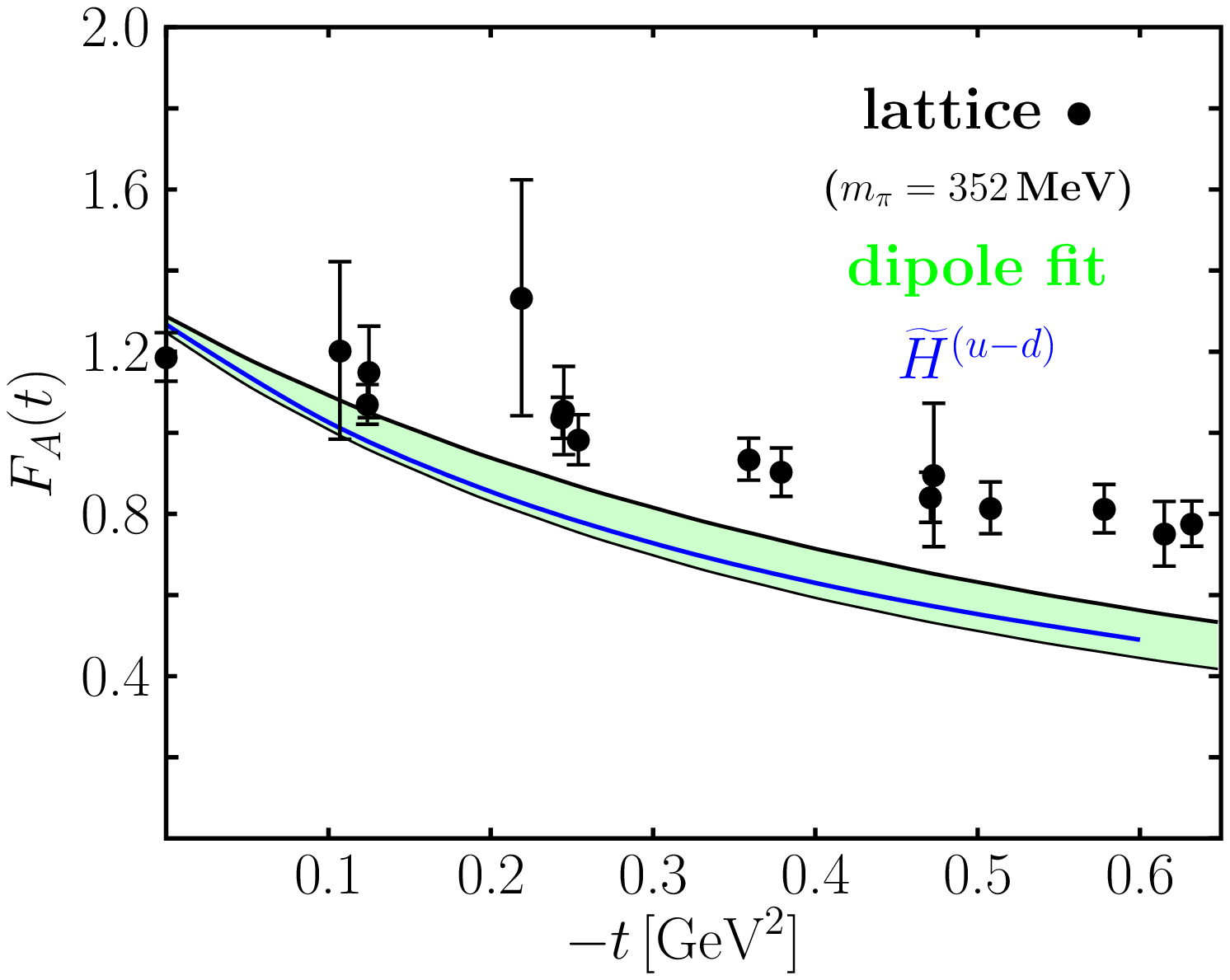}
\includegraphics[width=0.45\tw,bb=158 370 589 720,clip=true]{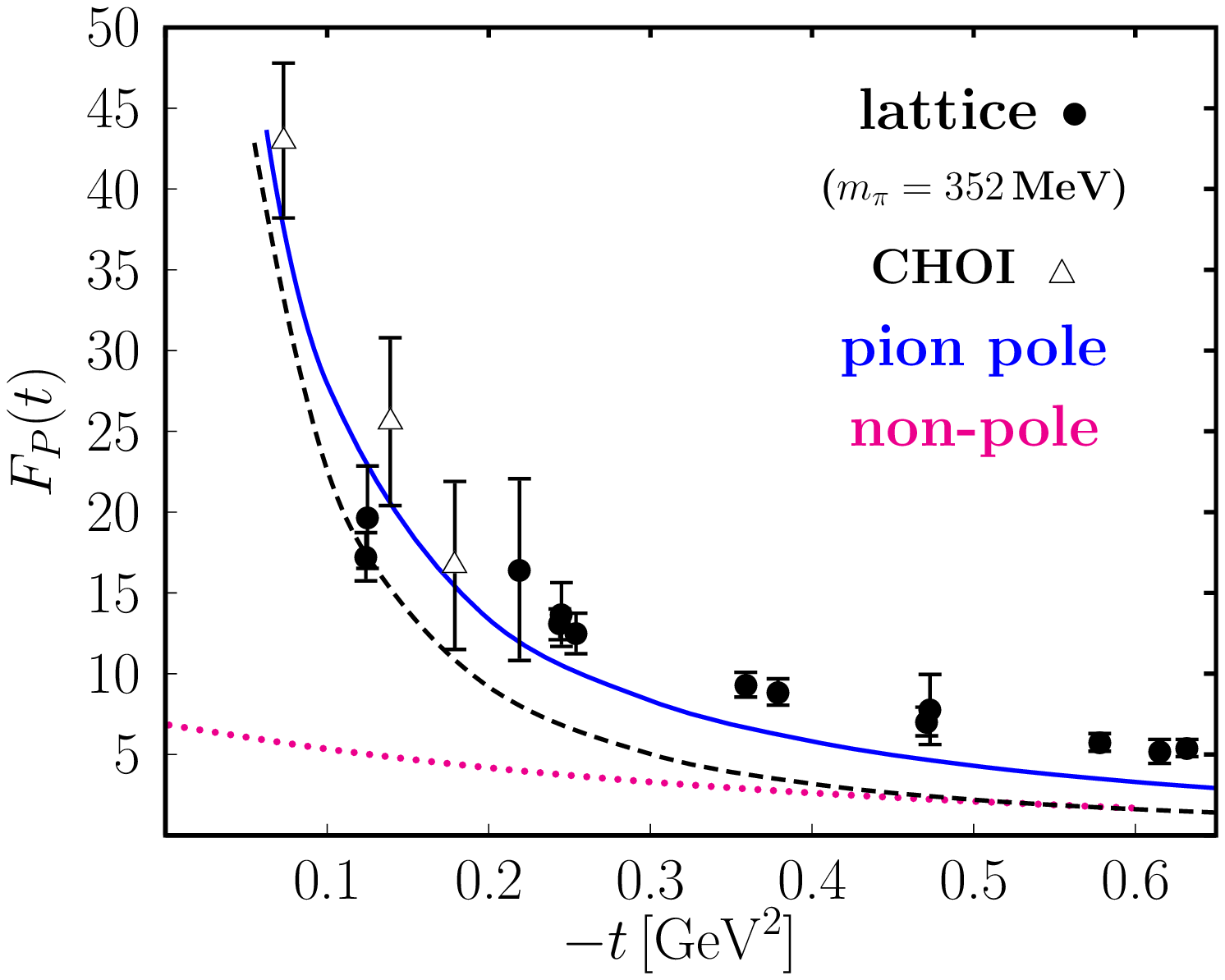}
\caption{\label{fig:FF} The axial (left) and the pseudoscalar (right) nucleon 
form factors versus $-t$. The solid circles represent the lattice results 
\ci{lattice07} for a pion mass of $352\,\mev$. The experimental data are taken 
from \ci{kitagaki} (presented as a dipole fits with the errors shown as the 
green band) and from \ci{choi} (open triangles). The form factors evaluated
from our GPDs are shown as solid lines. The dashed and dotted lines represent 
the pion-pole (with $\Lambda_N=0.44\,\gev$) and non-pole contributions to
$F_P$, respectively. (Color online)}
\end{center}
\end{figure} 

As a first check of our parameterization we evaluate the pseudoscalar form
factor for the $p\to n$ transition, traditionally denoted by $F_P$
\be
F_P(t)\=F_P^{\rm pole}(t) +\int_{-1}^1 d\xb\,
                          \widetilde{E}_{\rm n.p.}^{(3)}(\xb,\xi,t)\,.
\ee
With the help of PCAC and the Goldberger-Treiman relation the pion-pole
contribution to the pseudoscalar form factor can be expressed by
\be
F_P^{\rm pole}(t)\=2mf_\pi\,\frac{\sqrt{2} g_{\pi pn} F_{\pi
    pn}(t)}{m_\pi^2-t}
\ee
where $f_\pi (=131\,\mev)$ is the pion decay constant , $g_{\pi pn} (=13.1)$
the pion-nucleon coupling constant \ci{hanhart} and
\be
F_{\pi pn}\=\frac{\Lambda_N^2-m_\pi^2}{\Lambda_N^2\, -\, t}\,.
\label{eq:Fpipn}
\ee
The results for $F_P(t)$ are shown in Fig.\ \ref{fig:FF} and compared to the
experimental data \ci{choi} and results from lattice QCD \ci{lattice07}. One
notes the same features as for the axial form factor - reasonable agreement at 
low $-t$ and a flatter $t$ dependence of the lattice results. The dominance 
of the pion pole at small $t$ is clearly visible.

From the analysis of $\pi^+$ electroproduction \ci{GK5} only the isovector
combination of $\widetilde{E}_{\rm n.p.}$ is fixed. In the absence of any
experimental information that would allow to fix the normalization of the 
isoscalar combination too we have to rely on lattice-QCD results. 
Adjusting the normalizations to the second moments of $\widetilde{E}$ given 
in \ci{lattice07} which  are free from the pion-pole contribution, we arrive
at the values of the normalization constants for individual flavors quoted 
in Tab.\ \ref{tab:1}. The lattice results \ci{lattice07} are however not 
accurate enough to allow for a determination of further differences in the 
parameterization \req{eq:pdf} of $\widetilde{E}_{\rm n.p.}$ for $u$ and $d$ 
quarks and, we repeat, are not extrapolated to the chiral limit.  
  
\subsection{$H_T$}
The forward limit of $H_T$, the transversity PDF, is taken from an updated
version of the analysis of the data on the azimuthal asymmetry in
semi-inclusive deep inelastic scattering and in inclusive two-hadron
production in electron-positron annihilation \ci{anselmino08}. Since the 
parameters quoted in \ci{anselmino08} differ from those obtained in
a previous anlaysis which has been used by us in \ci{GK5}, our parameterization
of $H_T$ is to be changed accordingly. Now, we parameterize the transversity
PDFs as
\be
\delta^a\=N^a_{H_T}\, \rho^{1/2} (1-\rho)\big[q^a(\rho)+\Delta q^a(\rho)\big]\,.
\label{eq:transversity-pdf}
\ee
This parameterization corresponds to a Regge intercept of about zero.
The PDFs are taken from \ci{BB,cteq6}. The GPD is then calculated from
\req{eq:DD} and \req{eq:integral} with the Regge parameters quoted in Tab.\
\ref{tab:1}.

\begin{figure}[t]
\begin{center}
\includegraphics[width=0.45\tw,bb=138 350 589 720,clip=true]
{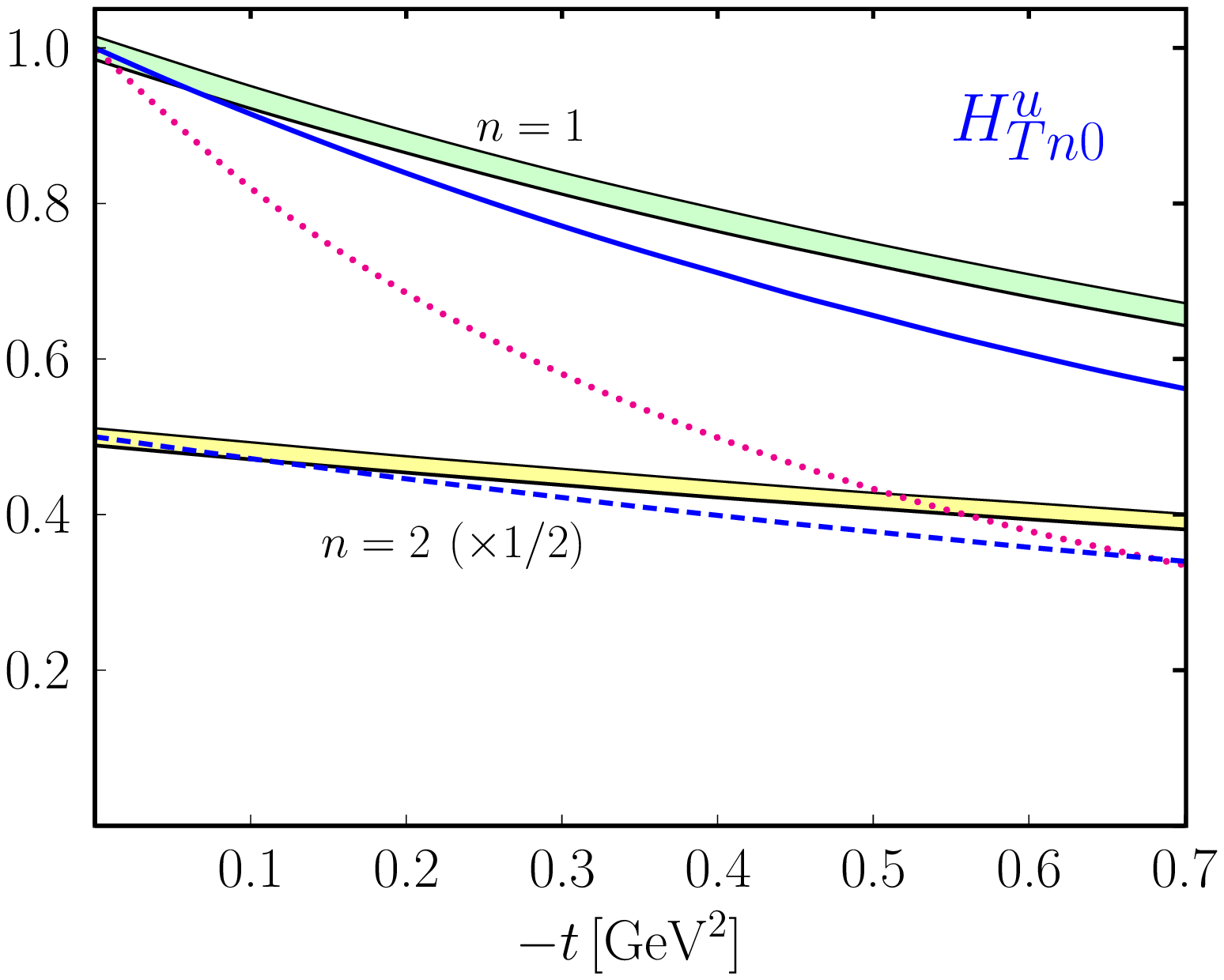}
\includegraphics[width=0.45\tw,bb=138 350 589 720,clip=true]{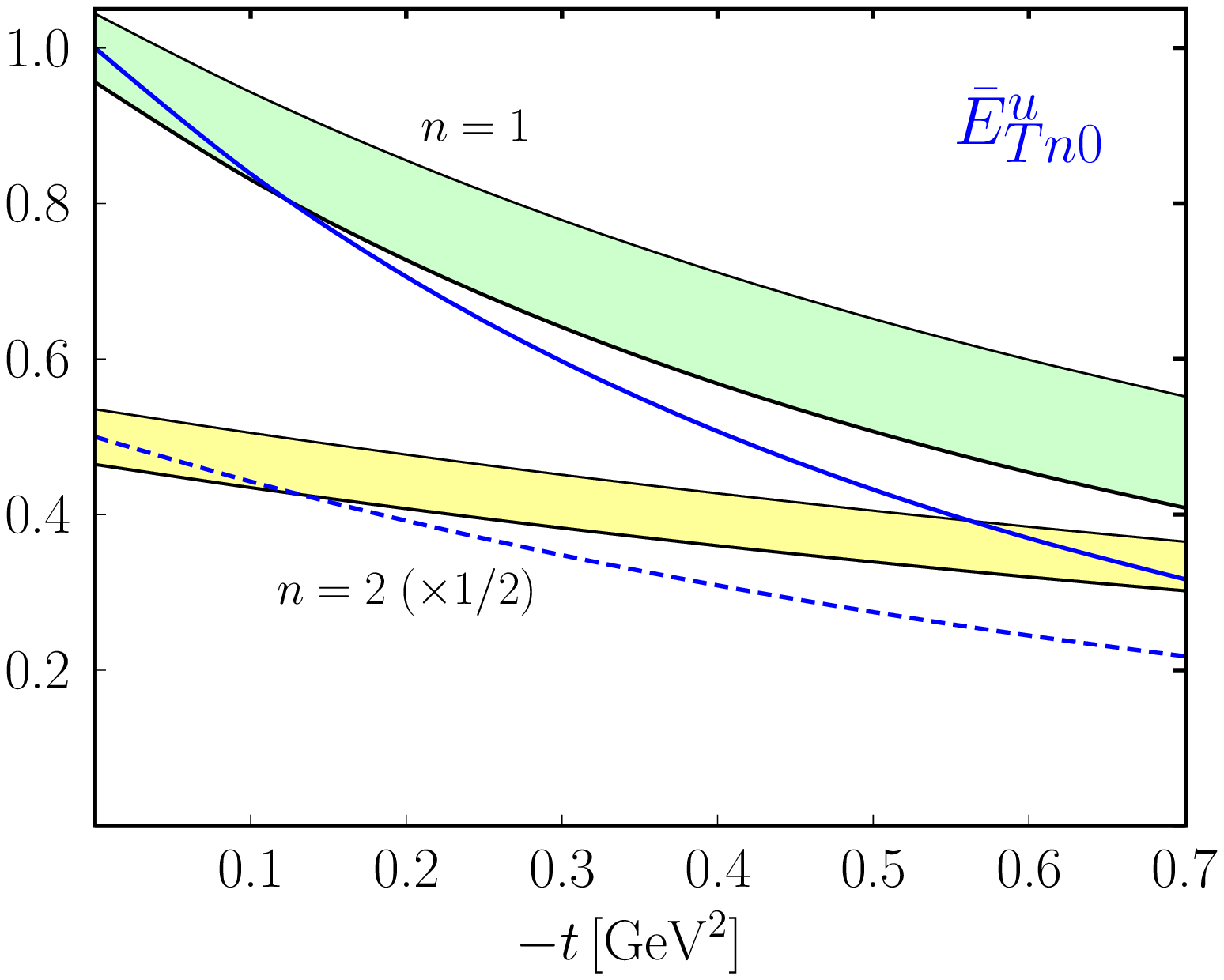}
\caption{\label{fig:HT-ET} The first and second moments of $H_T$ (left) and
  $\bar{E}_T$ (right) for $u$ quarks scaled by the forward value. The results 
  evaluated from our GPDs are shown as solid and dashed lines, the lattice-QCD 
  results \ci{goeckeler,lattice05} as bands indicating their
  uncertainties. The dotted line represents the chiral quark-soliton model 
  \ci{ledwig}. (Color online)}
\end{center}
\end{figure}

Moments of $H_T$ have been calculated in \ci{lattice05}. They are represented
as dipole fits with extrapolations to the physical pion mass. In Tab.\
\ref{tab:2} the two lowest moments at $t=0$ are compiled and their $t$ 
dependence is shown and compared to the moments evaluated from our GPDs in 
Fig.\ \ref{fig:HT-ET}. Our moments at $t=0$ which are practically those
obtained by Anselmino {\it et al} \ci{anselmino08}, are about $40\%$ smaller
than the lattice results. As for the other GPDs our moments possess a somewhat
steeper gradient than the lattice moments as, at least for $u$
quarks, can be seen from Fig.\ \ref{fig:HT-ET}. The first moment of $H_T$,
termed the tensor form factor,
\be
g_T^a(t) \= \int_{-1}^1 d\xb\, H_T^a(\xb,\xi,t)\,, 
\ee 
has also been calculated within a chiral quark-soliton model \ci{ledwig}. At
$t=0$ the following values have been obtained in \ci{ledwig}
\be
H^u_{T10}(0)\= 0.876\,, \qquad H^d_{T10}(0) \= -0.251\,,
\ee
which are close the lattice results, and therefore larger than our ones,
too. On the other hand, the $t$ dependence obtained within that chiral
quark-soliton model is even steeper than we found, see Fig.\ \ref{fig:HT-ET}.  
Several other models (cf.\ \ci{anselmino08} and references therein) also 
predict larger moments of the transversity PDFs than those obtained from
the parameterization \req{eq:transversity-pdf} which is based on the analysis
performed in \ci{anselmino08}.

\begin{table*}[t]
\renewcommand{\arraystretch}{1.4} 
\begin{center}
\begin{tabular}{| c | c | c || c | c | c |}
\hline  
  & \ci{lattice05} &this work    &  &\ci{goeckeler} & this work \\[0.2em]
\hline
$H^u_{T10}$ & $\phantom{-}0.857 (13)$ & $\phantom{-}0.585$ & $\bar{E}^u_{T10}$ 
& $2.93 (13)$ & $2.93$ \\[0.2em]
$H^u_{T20}$ & $\phantom{-}0.268 (6)$ & $\phantom{-}0.123$ & $\bar{E}^u_{T20}$ 
& $0.420 (31)$ & $0.360$\\[0.2em]
$H^d_{T10}$ & $-0.212 (5)$ & $-0.153$ & $\bar{E}^d_{T10}$  
& $1.90 (9)$  &  $1.90$\\[0.2em]
$H^d_{T20}$ & $-0.052 (2)$ & $-0.021$ & $\bar{E}^d_{T20}$  
& $0.260 (23)$  & $0.199$ \\[0.2em]
\hline
\end{tabular}
\end{center}
\caption{The first two moments of the transversity PDFs $H_T$ and 
$\bar{E}_T$ at $t=0$ quoted at the scale $2\,\gev$.}
\label{tab:2}
\renewcommand{\arraystretch}{1.0}   
\end{table*}

\subsection{$\bar{E}_T$}
The GPD $\bar{E}_T$ defined in \req{eq:combination}, is a new element in our
analysis, it has been ignored in \ci{GK5}. According to lattice QCD
\ci{goeckeler} $\bar{E}_T$ has however large moments with the same sign and
similar size for $u$ and $d$ quarks. It therefore seems to be expedient to 
examine its impact on electroproduction of pseudoscalar mesons.

The GPD $\bar{E}_T$ is parameterized as the other ones, \req{eq:DD} -  
\req{eq:integral}, but in contrast to the situation for the other GPDs, its
parameters can only be fixed with the help of lattice results. The latter are
presented in \ci{goeckeler} in form of a $p$-pole fit
\be
 \bar{E}^a_{Tn0}(t)\= \frac{\bar{E}^a_{Tn0}(0)}{\big[1-t/(p m^2_{an})^2\big]^p}
\ee
with $p=2.5$. The forward values, quoted in Tab.\ \ref{tab:2}, are extrapolated
to the chiral limit but not the mass parameters. The lowest moments at 
$t=0$ are the analogues of the familiar anomalous magnetic moments, $\kappa^a$,
and are therefore termed tensor anomalous magnetic moments, $\kappa^a_T$. The 
lattice results for them are of the same magnitude as the usual ones but in
contrast to those they are both positive. This is in agreement with model
studies performed in \ci{burkhardt06} which also support the expectation that 
$\kappa^a_T \gsim \mid\kappa^a\mid$. The lattice result is further 
corroborated by large-$N_c$ considerations \ci{burkhardt06}  which provide
$\kappa^u_T\approx \kappa^d_{T}$ and by a recent constituent quark model
\ci{pasquini}. It has been speculated \ci{burkhardt06} that $\bar{E}_T$
is linearly related to the Boer-Mulders function, $h_1^{\bot a}$,
\ci{boer-mulders} while the Sivers function, $f_{1\top}^a$,  is analogously
related to the transversity PDFs. In a recent analysis \ci{prokudin} of the 
$\langle \cos{(2\phi)}\rangle$ measurements \ci{compass,hermes} some evidence
has been obtained that the signs of the Boer-Mulders and the Sivers functions 
exhibit  the same pattern as those of the lowest moments of $\bar{E}_T$ and 
$H_T$ at $t=0$ which are quoted in Tab.\ \ref{tab:2}.  

The parameters of $\bar{E}_T$ are fixed by us in such a way that the lowest
moments from lattice QCD at $t=0$ are exactly reproduced and the second 
moments at least approximately. The $t$ dependence is assumed to be somewhat 
steeper than the lattice results as is the case for the other GPDs. This can 
be achieved with the parameters 
\be
\beta^u_{E_T}\=4\,, \qquad \beta^d_{E_T}\=5\,.
\ee
and those quoted in Tab.\ \ref{tab:1}. The moments at $t=0$ are given in Tab.\
\ref{tab:2} and their $t$ dependencies are shown in Fig.\ \ref{fig:HT-ET}.

We have checked that our GPDs respect various positivity bounds. Thus, for
instance,
\be
(E\pm \bar{E}_T)^2 (t=0) \leq 4m^2 (H-\widetilde{H})\frac{\partial}{\partial
  t}(H+\widetilde{H}\pm 2H_T) (t=0) \,,
\ee
and similar ones derived in \ci{diehl-haegler} at zero skewness or bounds on 
$\widetilde{H}$ and $\widetilde{E}$ in  terms of PDFs \ci{poby} as for instance
\ba
\left|\frac{\sqrt{-t^\prime}}{2m}\,\xi\widetilde{E}^q\right| &\leq&
\frac14 \Big[\sqrt{(q+\Delta q+2\delta)_{x}\,(q-\Delta q)_{x^\prime}} \nn\\ 
      &+&\sqrt{(q-\Delta q)_{x}\,(q+\Delta q+2\delta)_{x^\prime}}\nn\\
       &+&\sqrt{(q+\Delta q-2\delta)_{x}\,(q-\Delta q)_{x^\prime}}\nn\\ 
        &+&\sqrt{(q-\Delta q)_{x}\,(q+\Delta q-2\delta)_{x^\prime}}\,\Big]\,, 
\label{eq:poby}
\ea
where the GPDs are to be taken at $\xb$, $\xi$ and $t$ while the PDFs are
either to be evaluated at $x$ or $x^\prime$, the individual momentum fractions
of the emitted or reabsorbed partons,
\be
x \= \frac{\xb + \xi}{1+\xi}\,, \qquad  x^\prime \= \frac{\xb-\xi}{1-\xi}\,.
\ee
All bounds hold for a given flavor; \req{eq:poby} only holds in the DGLAP
region. Flavor labels are omitted for convenience in \req{eq:poby}.

\section{Electroproduction of pions}
\label{sec:pion}

\begin{figure}[t]
\begin{center}
\includegraphics[width=0.50\tw]{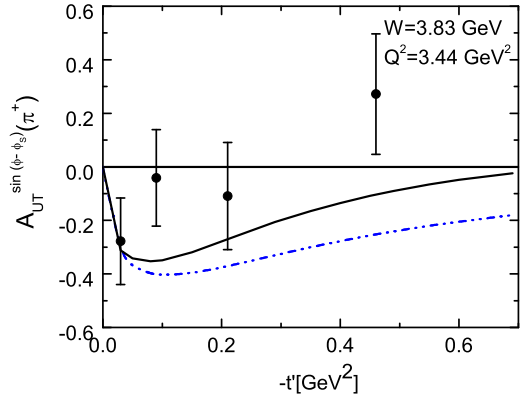}
\caption{Results for the  $\sin{(\phi-\phi_s)}$ moment for $\pi^+$ production
  shown as solid line. The dash-dot-dotted line is obtained under neglect of 
  $\bar{E}_T$. Data are taken from \ci{Hristova}. (Color online)}  
\label{fig:aut-pi+}
\end{center}
\end{figure}

Since we have slightly changed the parameterizations of the GPDs as compared 
to \ci{GK5} and have taken into account $\bar{E}_T$ now, we have to repeat 
the computation of $\pi^+$ electroproduction in order to examine the quality 
of the new results. As a matter of fact, it turns out that they are very close 
to the previous results presented in \ci{GK5}. Their agreement with experiment
\ci{hermes07,Hristova,hermes02} is reasonable. The contribution from
$\bar{E}_T$ to the cross section is tiny, much smaller than the errors of the 
HERMES cross section data \ci{hermes07}, since the flavor combination 
$\bar{E}^u_T-\bar{E}^d_T$, occuring in $\pi^+$ electroproduction, is small. It
is only noticeable in the asymmetry parameters where it improves the agreement 
with experiment in general. As an example the $\sin{(\phi-\phi_s)}$ moment of 
the cross section measured with a transversely polarized target \ci{Hristova}
is shown in Fig.\ \ref{fig:aut-pi+} ($\phi$ is the azimuthal angle between the 
lepton and the hadron plane while $\phi_s$ specifies the orientation of the 
target spin vector with respect to the lepton plane). 

\begin{figure}[t]
\begin{center}
\includegraphics[width=0.45\tw]{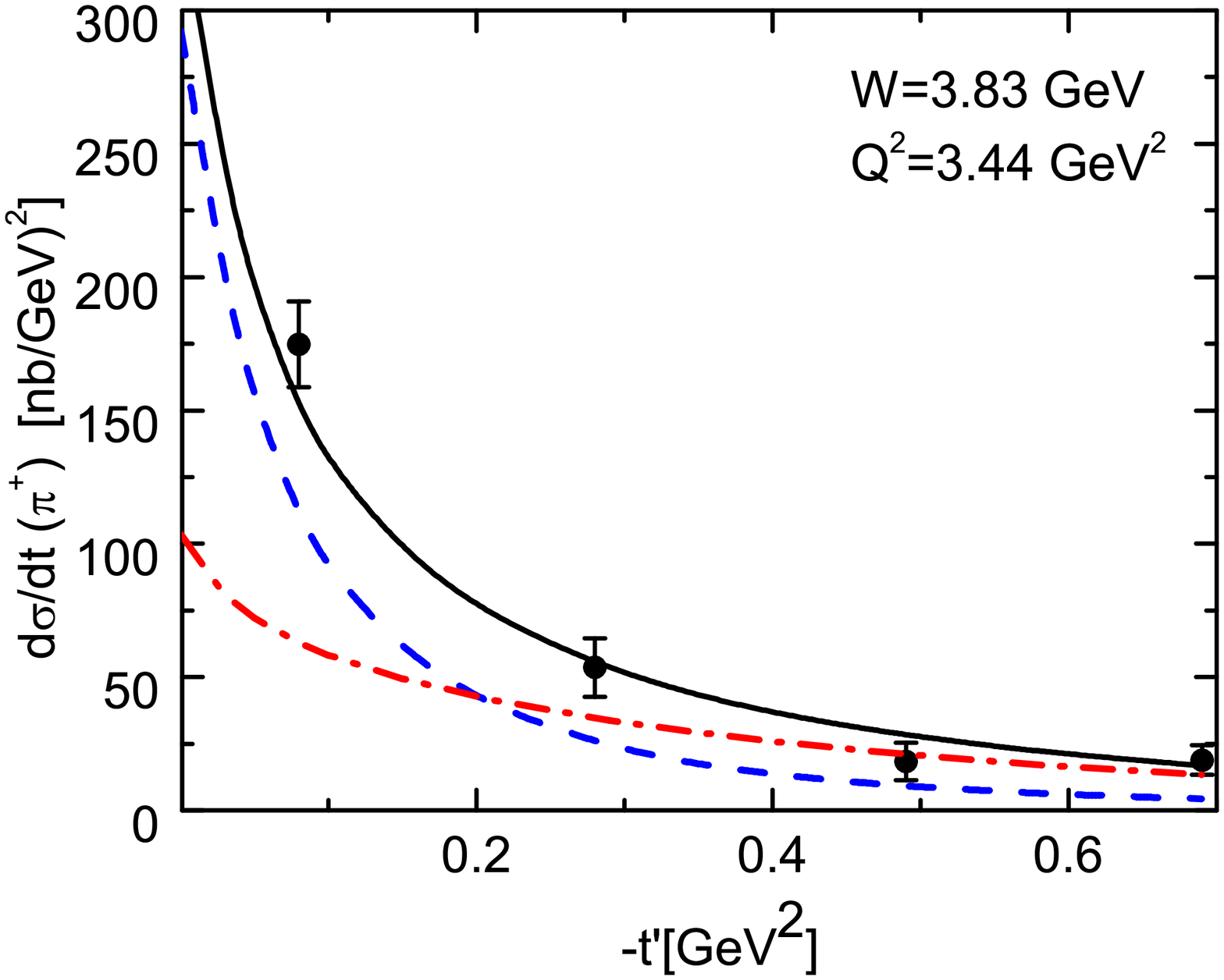} 
\includegraphics[width=0.45\tw]{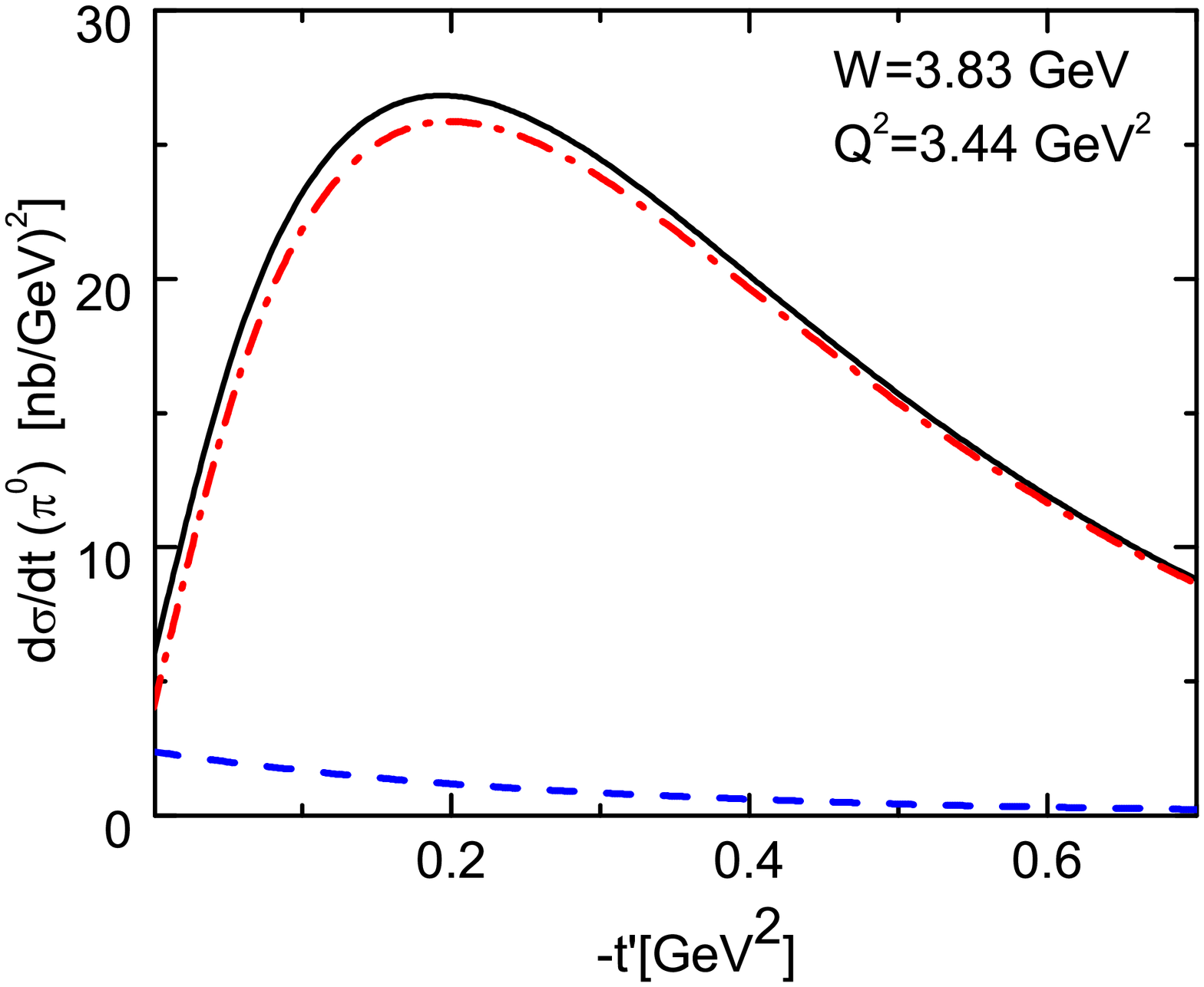}
\caption{Results for the $\pi^+$ (left) and $\pi^0$ (right) cross sections. 
  The solid (dashed, dash-dotted) lines represent the unseparated
  (longitudinal, transverse) cross sections. The $\pi^+$ data are taken from 
  \ci{hermes07}. (Color online)}  
\label{fig:sigma-pi+-pi0}
\end{center}
\end{figure}

As a second example the unseparated cross section as well as its longitudinal
and transverse components
 ($d\sigma=d\sigma_T + \varepsilon d\sigma_L$) for $\pi^+$ production are
 shown in Fig.\ \ref{fig:sigma-pi+-pi0} at $Q^2=3.44\,\gev^2$ and 
$W=3.83\,\gev$. For the ratio of the longitudinal and transverse photon
fluxes, $\varepsilon$, a value of $0.8$ is taken. Evidently, the agreement
with the HERMES data \ci{hermes07} is very good. The longitudinal cross
section is dominant at low $-t^\prime$ but if $-t^\prime$ becomes larger than 
about $0.2\,\gev^2$ the transverse cross section takes the lead. A crossing 
of the two cross sections is also seen in the large-skewness Jefferson Lab
data measured by the $F_\pi$ collaboration \ci{blok} although it occurs at a 
smaller value of $-t^\prime$. Actually the crossing takes place at about the 
same value of $t$ for our results and the $F_\pi$ data. It is to be stressed 
that the dominance of the longitudinal cross section at small $-t^\prime$ is a
consequence of the strong pion pole. Without it one would have
$d\sigma_T>d\sigma_L$ at all $t^\prime$. The pure pole contribution to
the longitudinal cross section behaves as $-t/(t-m_\pi^2)^2$ at small $-t$.
This factor has a pronounced maximum at $t=-m_\pi^2$. For $\xi\geq m_\pi/(2m)$ 
the maximum lies outside the scattering region. Hence, the longitudinal cross 
section decreases continuously. On the other hand, for $\xi<m_\pi/(2m)$ the
above factor generates a maximum of $d\sigma_L/dt$ at a small value of 
$-t^\prime$. The maximum is visible in the cross section if $Q^2\lsim (m_\pi/m) W^2$.

Before we turn to the discussion of $\pi^0$ production it is in order to
comment on the pion \wf s which are needed in the calculation of the subprocess
amplitudes (cf.\ \req{mod-amp}). In fact we use the same \wf s as in \ci{GK5},
namely a Gaussian in $\vk$ with an associated distribution amplitudes
$\Phi_{AS}=6\tau (1-\tau)$ for the twist-2 subprocess amplitudes and
$\Phi_p\equiv 1$ for twist 3. For the transverse-size parameter in the
Gaussian we use $a_\pi=[\sqrt{8}\pi f_\pi]^{-1}$ ($=0.861\,\gev^{-1}$) for
twist 2 and $a_{\pi  p}=1.8\,\gev^{-1}$ for twist 3. There is an extra factor 
$\vk$ multiplying the Gaussian for the twist-3 \wf. For a detailed discussion 
of such a $\vk$ factor and its interpretation see \ci{ji-yuan,bolz94}. 
Higher-order Gegenbauer terms in the distribution amplitudes are strongly 
suppressed in the modified perturbative approach: The Sudakov factor in 
conjunction with the subprocess amplitude provides a series of power 
suppressed terms which are generated in the region of soft quark momenta 
($\tau, 1-\tau\to 1$). They grow with the Gegenbauer index and reduce the 
perturbative contribution. With increasing $Q^2$ the higher Gegenbauer terms 
become gradually more important (cf.\ the discussion in \ci{kroll10}). Since 
we are merely interested in rather small values of $Q^2$ the asymptotic 
distribution amplitude suffices for pion production.

Now, for $\pi^0$ electroproduction there is no contribution from the pion pole
and the GPDs appear in the flavor combination
\be 
(e_uF_i^u-e_dF_i^d)/\sqrt{2} 
\label{eq:flavor-pi0}
\ee
in contrast to $F_i^u-F_i^d$ for $\pi^+$ production~\footnote{
For the production of $\pi^+$ and kaons the quark charges are absorbed in the
hard scattering amplitude.}.
From the relative signs and sizes of the GPDs for $u$ and $d$ quarks it is
evident that the contributions from $\widetilde{H}$ and $H_T$ are large for
$\pi^+$ but small for $\pi^0$ production while for $\widetilde{E}_{\rm n.p.}$
and $\bar{E}_T$ the situation is reversed. The contributions from the latter
two GPDs are small for $\pi^+$ but large for $\pi^0$ production. As a
consequence the longitudinal $\pi^0$ cross section is much smaller than the 
corresponding $\pi^+$ one even if the pion-pole contribution is subtracted. 
Another consequence is that $\pi^0$ production is dominated by contributions 
from transversely polarized photons which are mainly generated by $\bar{E}_T$. 
The latter contribution is by order of magnitude stronger in $\pi^0$ than in 
$\pi^+$ production where a strong cancellation between $\bar{E}_T^u$ and 
$\bar{E}_T^d$ occurs. The importance of $\bar{E}_T$ in $\pi^0$ production is 
obvious from  the forward dip in the cross section (see Fig.\ 
\ref{fig:sigma-pi+-pi0}); it only contributes to helicity flip amplitudes, see 
\req{eq:ebar-ampl}. 
\begin{figure}[t]
\begin{center}
\includegraphics[width=0.45\tw]{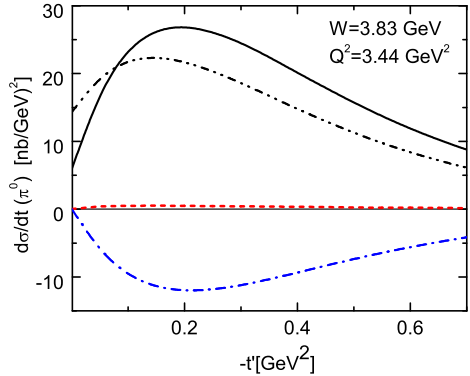} 
\caption{Unseparated and partial cross sections for $\pi^0$ production. 
  The solid (dashed, dash-dotted) line represents the unseparated 
  (longitudinal-transverse, transverse-transverse) cross section. The 
  dash-dot-dotted line is evaluated from the modified GPDs, see text. 
  (Color online)}  
\label{fig:sigma-pi0-lttt}
\end{center}
\end{figure} 

\begin{figure}[t]
\begin{center}
\includegraphics[width=0.45\tw]{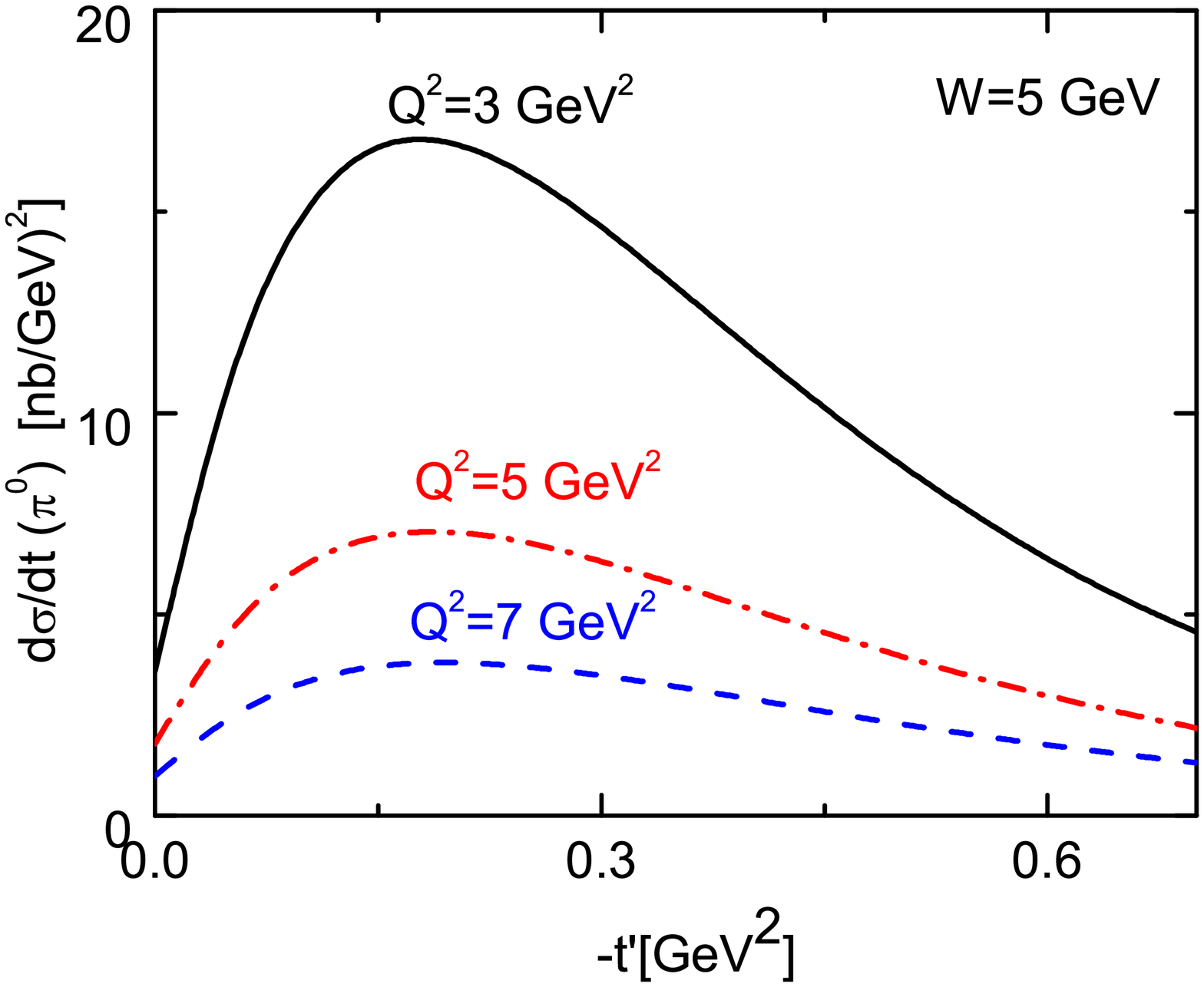} 
\includegraphics[width=0.45\tw]{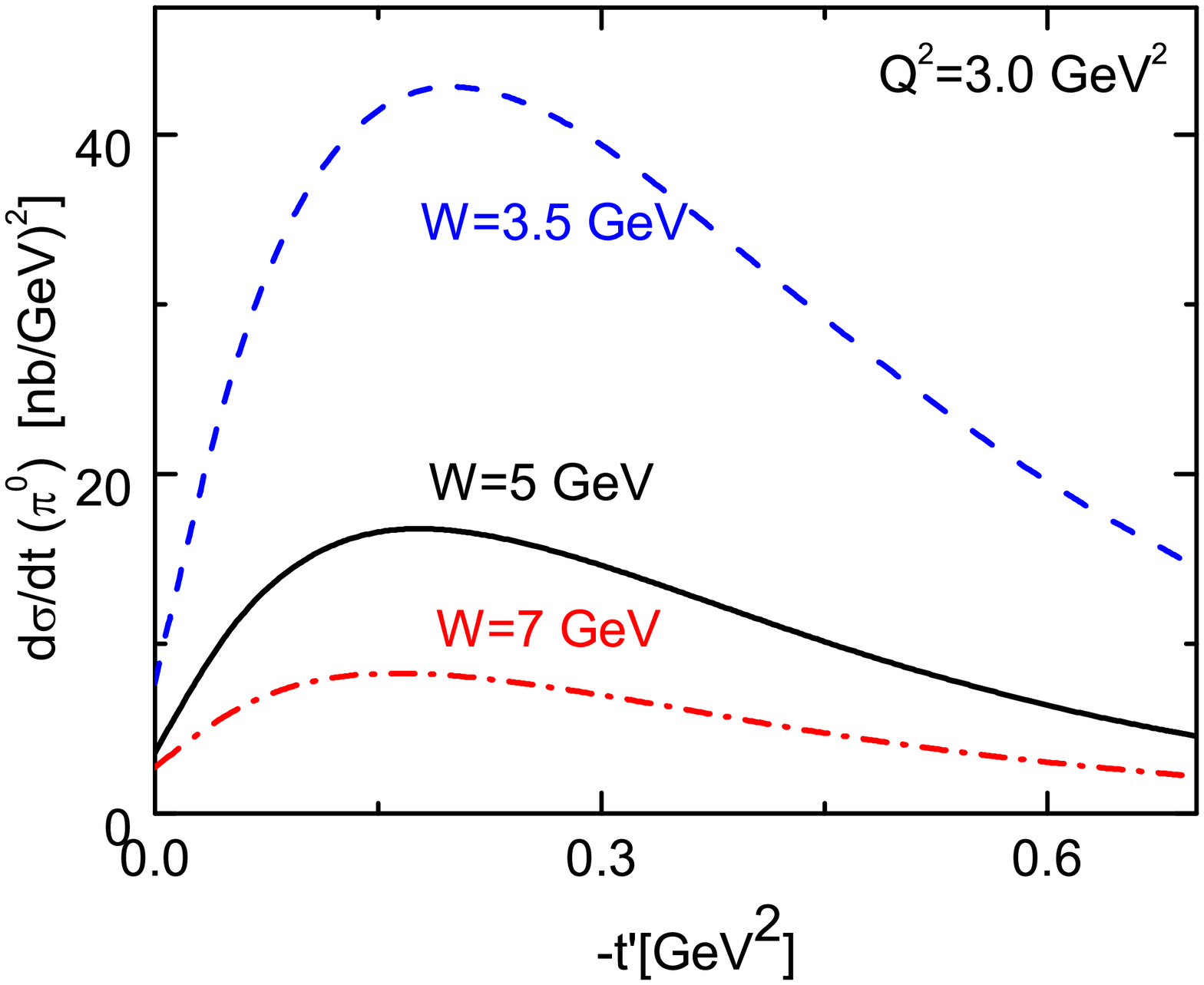}
\caption{Results for $\pi^0$ the cross section versus $t^\prime$ at
  $W=5\,\gev$ for sample values of $Q^2$ (left) and at $Q^2=3\,\gev^2$ for
  various values of $W$ (right). The polarization of the photon flux is 
  taken as $\varepsilon=0.35$ at $W=3.5\,\gev$ and $0.8$ for
  $W\geq 5\,\gev$. (Color online)}  
\label{fig:sigma-pi0}
\end{center}
\end{figure}

As we repeatedly mentioned our GPDs are optimized for the range of small
skewness. Therefore, our approach cannot readily be applied at the kinematics 
accessible at the Jefferson Lab. Ignoring this supposition and applying our 
approach in its present form at Jefferson Lab kinematics straightforwardly, 
one runs into difficulties with $\rho^0$ production - the cross section is at 
variance with experiment \ci{clas-rho} by order of magnitude
\ci{GK2,kroll:10}. For $\phi$ production \ci{clas-phi}, on the other hand, our 
approach works quite well. The reason for this failure of the handbag approach 
with $\rho^0$ production is not understood but it is likely be related to the 
valence quarks. Similar difficulties may happen for electroproduction of 
pions. With this admonition in mind one may compare our approach with the 
preliminary CLAS data on $\pi^0$ production \ci{kubarowsky} 
(at $W\simeq 2.5\,\gev$ and $Q^2\simeq 2.3\,\gev^2$). It turns out that the 
magnitude of the cross section is  about right. Even a little forward dip 
is seen in the CLAS data on the cross section. This effect has also been
observed in a Hall A experiment \ci{halla}. It is important to realize that 
the depth of the dip is determined by the ratio of the amplitudes 
${\cal M}_{0-,++}$ and ${\cal M}_{0+,+ +}$, i.e.\ it is influenced by the
relative strength of $H_T$ and $\bar{E}_T$. We remind the reader that the 
normalization of the first GPD is fixed by the transversity PDF determined in 
\ci{anselmino08} while the second one is constrained by lattice-QCD results 
\ci{goeckeler}. A change of these GPDs may lead to a different shape of the 
cross section in the forward region. An example of such a modification of the 
GPDs is also shown in Fig.\ \ref{fig:sigma-pi0-lttt} and compared to the result 
evaluated from the standard parameterization of the GPDs presented in Sect.\ 
\ref{sec:GPD}. For this variant we modify $H_T$ by changing its normalization: 
$N^u_{H_T}=1.1$ and $N^d_{H_T}=-0.3$ and take for the slope $b_{H_T}=0.3\,\gev^{-2}$
(see Tab.\ \ref{tab:1}). In order to have more or less the same $\pi^0$ cross
section at CLAS kinematics the GPD $\bar{E}^a_T$ is multiplied by a factor
0.8. For this modification of the transversity GPDs our fit to $\pi^+$
production cross section practically remains unchanged. It is of the same
quality as for the standard parameterization since $H^u_T-H^d_T$ is almost 
unaltered. On the other hand, the forward limits of $H_T$ are now at variance 
with the transversity PDFs given in \ci{anselmino08} and the lowest moments of 
$\bar{E}_T$ are below the lattice-QCD results \ci{goeckeler}. As one sees from 
Fig.\ \ref{fig:sigma-pi0-lttt} the dip at $t^\prime$ is considerably less deep
now. The longitudinal-transverse and transverse-transverse interference cross
sections are also shown for the standard parameterization  in this figure. The 
analogous predictions for CLAS kinematics are in fair agreement with
experiment \ci{kubarowsky}. The energy and $Q^2$ dependencies of the $\pi^0$
cross section are shown in Figs.\ \ref{fig:sigma-pi0}.

\begin{figure}[t]
\begin{center}
\includegraphics[width=0.45\tw]{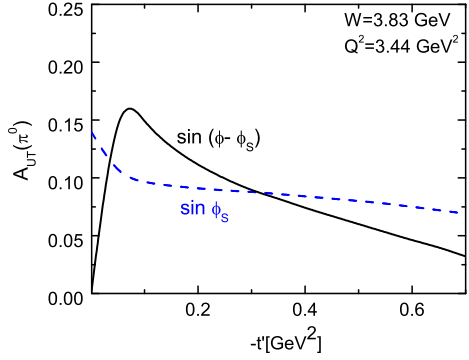} 
\caption{Results for the $\sin{(\phi-\phi_s)}$ and $\sin{\phi_s}$ moments of
  the transverse target asymmetries for $\pi^0$ electroproduction versus 
  $t^\prime$. (Color online)}
\label{fig:aut-pi0}
\end{center}
\end{figure}

Predictions for the $\pi^0$ target asymmetries are shown in Fig.\
\ref{fig:aut-pi0} for the standard kinematics as an example. The transverse 
target asymmetries behave differently for $\pi^0$ and $\pi^+$ production.
Particularly noteworthy is the fact the $\sin{\phi_s}$ moment for $\pi^0$ 
production is much smaller than for $\pi^+$ for non-zero $t^\prime$. This 
behavior can be traced back to the amplitude ${\cal M}_{0+,++}$ which is very 
small for $\pi^+$ but large for $\pi^0$ production. In terms of helicity 
amplitudes the asymmetry reads
\be
A_{UT}^{\sin{\phi_s}}\,\sigma_0 \simeq \sqrt{\varepsilon (1+\varepsilon)}
              {\rm Im}\Big[{\cal M}^*_{0+,++}\,{\cal M}_{0-,0+} - 
                 {\cal M}^*_{0-,++}\,{\cal M}_{0+,0+}\Big]\,,
\label{eq:sin-phi-s}
\ee
where it is assumed that the angle of the rotation in the lepton plane from the 
direction of the incoming lepton to the virtual photon is small. The quantity
$\sigma_0$ is given by
\ba
\sigma_0&=&\frac12 \Big[\mid{\cal M}_{0+,++}\mid^2+\mid{\cal M}_{0-,-+}\mid^2
     +\mid{\cal M}_{0-,++}\mid^2+\mid{\cal M}_{0+,-+}\mid^2\Big] \nn\\
 &+& \eps\; \Big[\mid{\cal M}_{0+,0+}\mid^2+\mid{\cal M}_{0-,0+}\mid^2\Big]\,.
\ea
The $\pi^+$ asymmetry is essentially generated by the second term in
\req{eq:sin-phi-s} while, for not too small $-t^\prime$, a strong cancellation
between both the terms in \req{eq:sin-phi-s} occur for $\pi^0$ production. For
$t^\prime\to 0$ the first term vanishes $\propto t^\prime$.

The different behavior of the $\sin{(\phi-\phi_s)}$ moment for $\pi^+$ and
$\pi^0$ production (see Figs.\ \ref{fig:aut-pi+} and \ref{fig:aut-pi0}) is
also a consequence of the contribution from $\bar{E}_T$ being small in one but
large in the other reaction. Approximately this asymmetry is given by
\be
 A_{UT}^{\sin{(\phi-\phi_s)}}\,\sigma_0 \simeq -2\varepsilon 
              {\rm Im}\Big[{\cal M}^*_{0-,0+}\,{\cal M}_{0+,0+}\Big] - 
              {\rm Im}\Big[{\cal M}^*_{0+,++}\,{\cal M}_{0-,++}\Big]
\label{eq:sin-phi-phis}
\ee
The second term provides a positive contribution to the $\pi^0$ asymmetry 
which overcompensates the asymptotically dominant first term provided by
longitudinally polarized photons. 

Our estimates of  $\pi^0$ electroproduction which, we repeat, strongly
depend on the lattice QCD findings \ci{goeckeler} for the moments of
$\bar{E}_T$, are distinct from our previous estimate presented in \ci{GK5} 
where the amplitude ${\cal M}_{0+,++}$ had been neglected, and from other
predictions. In collinear factorization to leading-twist accuracy only the 
longitudinal cross section can be calculated. The predictions for it, see 
e.g.\ \ci{goeke00}, are of about the same size as our longitudinal cross 
section but much smaller than the full cross section. In \ci{liuti08} the 
dominance of the transverse cross section in $\pi^0$ production has also 
been suggested. However, our treatment of the twist-3 contribution  differs 
from the approach advocated in \ci{liuti08} markedly. In the latter work the 
subprocess is viewed as form factors for photon-pion transitions under the 
action of vector and axial-vector currents. A proportionality between 
$\bar{E}_T$ and $H_T$ is assumed where the constants of proportionality are 
set by the transverse anomalous magnetic moments for which the values 0.6 
for $u$ and 0.3 for $d$ quarks are utilized. This parameterization of 
$\bar{E}_T$ is not supported by lattice QCD \ci{goeckeler} (see Tab.\ 
\ref{tab:2}) and differs from our parameterization drastically: $\bar{E}_T$ 
in \ci{liuti08} is much smaller than our one and has opposite signs for $u$ 
and $d$ quarks. As a consequence the absolute value of the amplitude 
${\cal M}_{0-,++}$ is much larger than that of ${\cal M}_{0+,+ +}$ as opposed 
to our results. The results for the $\pi^0$ cross section are therefore 
substantially smaller in \ci{liuti08}. Instead of a forward dip a forward 
maximum occurs.

\section{Electroproduction of of $\eta$ and $\eta^\prime$}
\label{sec:eta}
\begin{figure}[t]
\begin{center}
\includegraphics[width=0.45\tw,bb=121 515 357 660,clip=true]{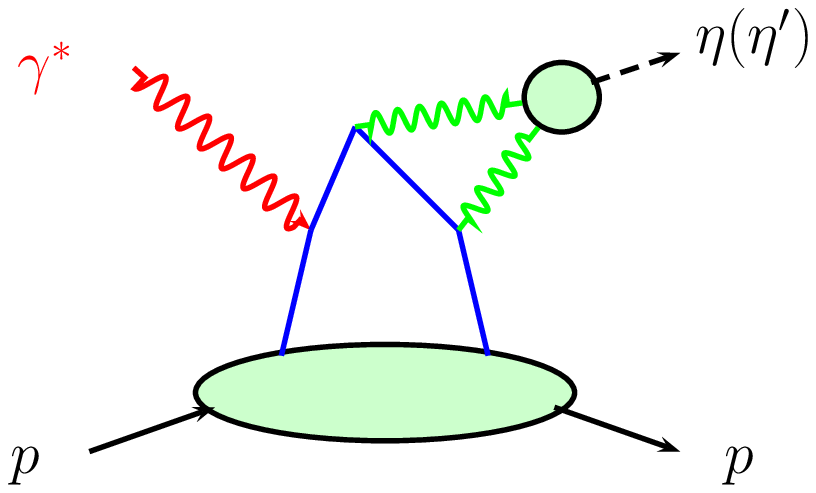}
\caption{\label{fig:gluon} A typical Feynman graph for the contribution from
  the gluon-gluon Fock component of the $\eta$ and $\eta^\prime$ mesons.}
\end{center}
\end{figure}

The treatment of the $\eta$ and $\eta^\prime$ mesons within the handbag
approach is similar to the case of the $\pi^0$. It is only a bit more 
intricate due to $\eta - \eta^\prime$ mixing and, in principle, the 
gluon-gluon Fock component of the $\eta$ and $\eta^\prime$ mesons. This 
Fock component contributes to the same order of $\als$ as the quark-antiquark 
components as can be seen from the Feynman graph shown in Fig.\ 
\ref{fig:gluon}. It has however been shown in \ci{passek} that the twist-2 
subprocess amplitude for the gluon-gluon component is $\propto t/Q^2$ and 
is therefore to be neglected for consistency. A possible twist-3 gluon 
contribution is neglected as well since, in contrast to the quark subprocess 
(cf.\ \req{eq:chiral}), there is no known reason why this contribution could 
be large. 

Working in the flavor octet-singlet basis and exploiting the results of the 
quark-flavor mixing scheme developed in \ci{FKS1}, we can decompose any of
the helicity amplitudes as 
\ba
{\cal M}^\eta &=& \cos{\theta_8}\,{\cal M}^{(8)} - \sin{\theta_1}\,
                {\cal M}^{(1)}\,,\nn\\
{\cal M}^{\eta^\prime} &=& \sin{\theta_8}\,{\cal M}^{(8)} + \cos{\theta_1}\,
                {\cal M}^{(1)}\,.
\ea
For the mixing angles we adopt the values 
\be
\theta_8\= -21.2^\circ\,, \qquad \theta_1\=-9.2^\circ
\ee
derived in \ci{FKS1} on exploiting the divergencies of the axial vector
current which embodies the axial-vector anomaly. The octet and singlet 
amplitudes are to be calculated in full analogy to the case of the $\pi^0$ 
with octet and singlet \wf s and decay constants taken from \ci{FKS1}
\be
f_8\= 1.26\,f_\pi\,, \qquad f_1\= 1.17\,f_\pi\,.
\label{eq:eta-decay-constants}
\ee
For these wave functions, assumed to be independent on the meson, we take
the same Gaussian as for the pion with exactly the same transverse size
parameters. For $Q^2$ in the range of interest in this work these simple \wf s
suffice, cf.\ the remark in Sect.\ \ref{sec:pion}. In the analysis of the
Babar data on the $P\gamma$ transition form factors \ci{babar}, also
performed within the modified perturbative approach \ci{kroll10}, only little 
differences between the octet and singlet \wf s have been found. The GPDs 
appear in the following flavor combinations
\ba
F_i^{(8)}&=&\frac1{\sqrt{6}}\,
    \big[e_u F_i^u + e_d F_i^d -2e_s F_i^s\big]\,, \nn\\
F_i^{(1)} &=& \frac1{\sqrt{3}}\,
    \big[e_u F_i^u + e_d F_i^d + e_s F_i^s\big]\,.
\label{eq:flavor-eta}
\ea
For the charge-conjugation-even mesons the GPDs contribute in the
valence-quark combination $F_i^a-F_i^{\bar{a}}$. For strange quarks we assume
$F_i^s\simeq F_i^{\bar{s}}$ for all GPDs~\footnote{
For the unpolarized strange quark PDFs \ci{cteq6s} and for $H^s$ \ci{diehl05} 
possible differences between strange and antistrange distributions have been 
studied. No evidence for a non-zero difference has been found within errors.}.
Hence, there is no contribution from strange quarks and one arrives at the
relation $F_i^{(1)}=\sqrt{2}F_i^{(8)}$.

\begin{figure}[t]
\begin{center}
\includegraphics[width=0.45\tw]{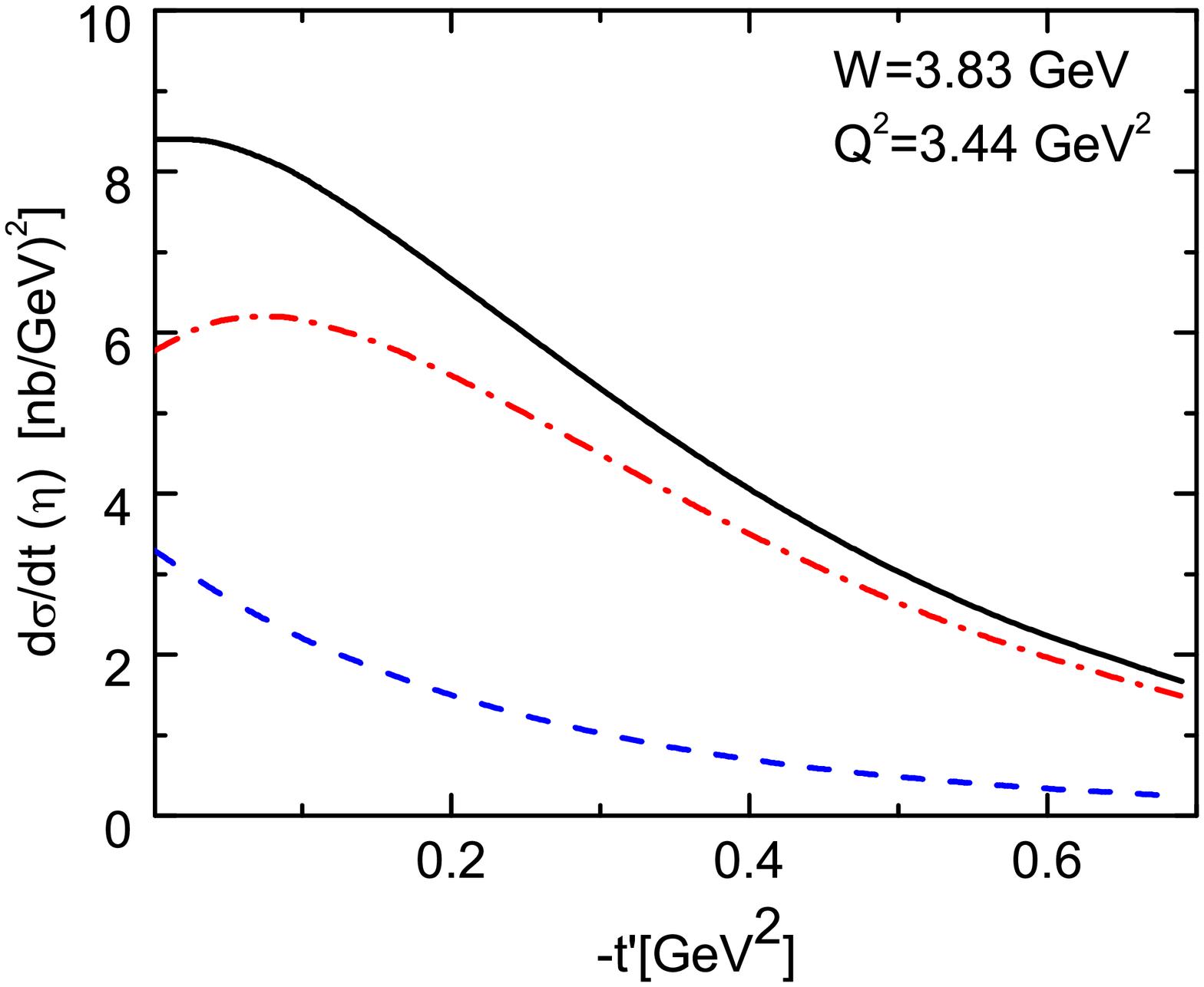} 
\includegraphics[width=0.45\tw]{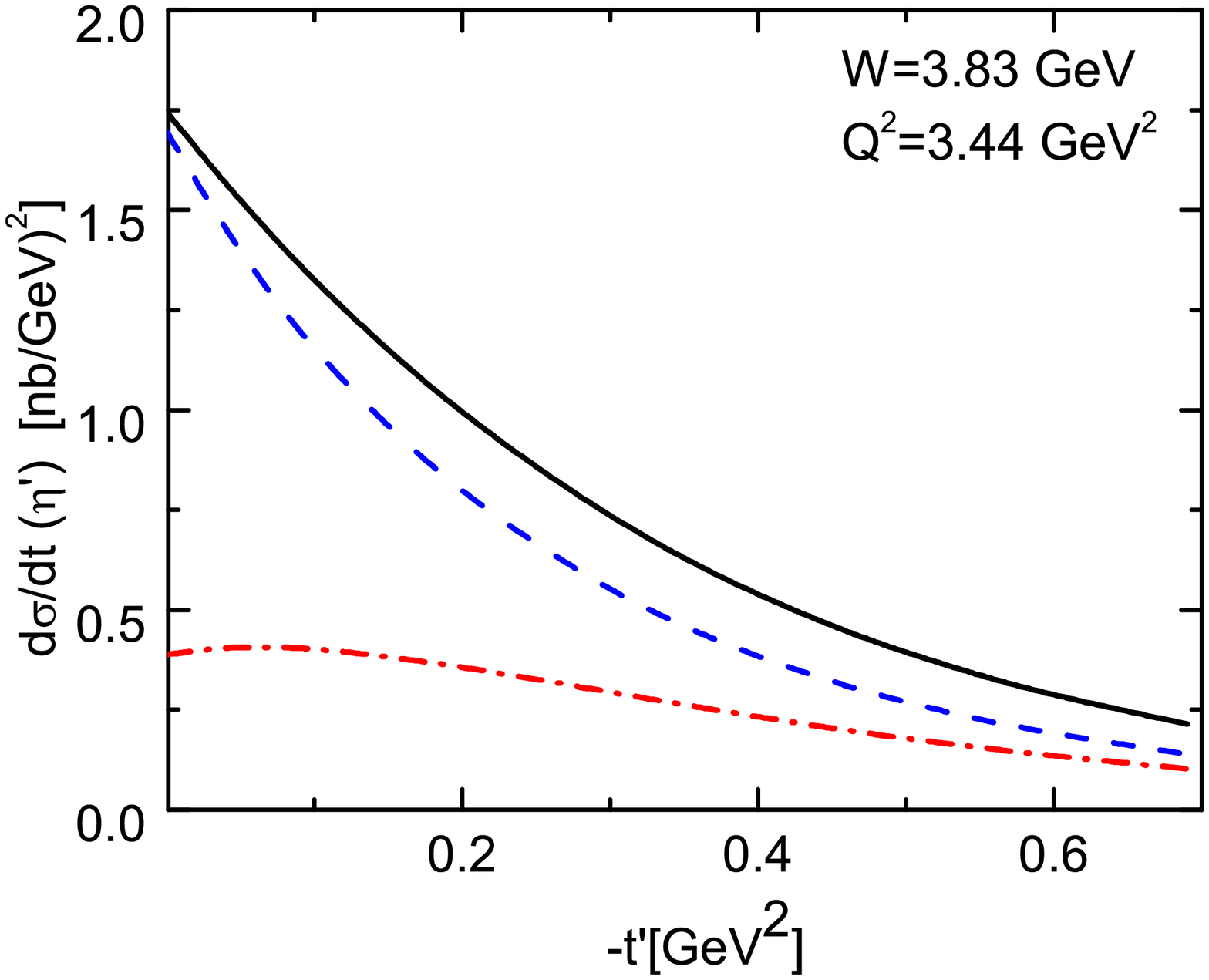}
\caption{Results for the $\eta$ (left) and $\eta^\prime$ (right) cross sections.
  For notations refer to Fig.\ \ref{fig:sigma-pi+-pi0}. (Color online)}  
\label{fig:sigma-eta-etap}
\end{center}
\end{figure}
\begin{figure}[t]
\begin{center}
\includegraphics[width=0.45\tw]{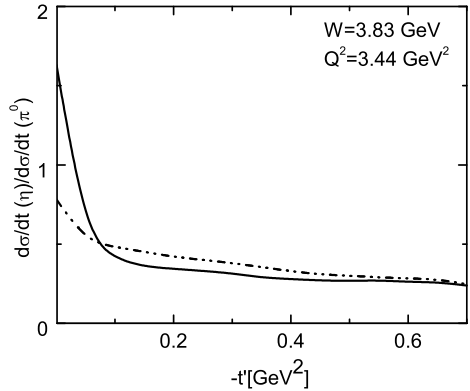} 
\caption{Ratio of the $\eta$ and $\pi^0$ cross sections. Solid
(dash-dot-dotted) line represents the results evaluated from the standard
parameterization of the GPDs (modified $H_T$). For notations refer 
to Fig.\  \ref{fig:sigma-pi+-pi0}. (Color online)}  
\label{fig:ratio-eta-pi0}
\end{center}
\end{figure}

 
The $\eta$ and $\eta^\prime$ cross sections are shown in Fig.\
\ref{fig:sigma-eta-etap}. In order to facilitate comparison of the
cross sections for the various mesons we use the same kinematical setting, 
$W=3.83\,\gev$ and $Q^2=3.44\,\gev^2$, for which there are $\pi^+$ data from 
HERMES \ci{hermes07}. For an understanding of the differences between the
$\eta$ and $\pi^0$ cross sections it is important to realize that the 
$u$ and $d$-quark GPDs contribute with opposite signs in these two reactions
(see \req{eq:flavor-eta} and \req{eq:flavor-pi0}). For small $-t^\prime$,
the region where the GPDs $\widetilde{H}$ and $H_T$ dominate, the opposite
sign of their $u$ and $d$-quark parts leads to a larger contribution to 
$\eta$ than to $\pi^0$ production which, in collaboration with the larger decay
constant \req{eq:eta-decay-constants}, overcompensates the relative factor
$1/\sqrt{3}$ between \req{eq:flavor-eta} and \req{eq:flavor-pi0}. Hence, 
the $\eta/\pi^0$ ratio of the cross sections is of order 1, see Fig.\ 
\ref{fig:ratio-eta-pi0}. For large $-t^\prime$ the GPD $\bar{E}_T$ 
dominates which has the same sign for $u$ and $d$ quarks. Therefore,  
the $\eta/\pi^0$ ratio is much smaller than 1; in fact close to $1/3$ 
for a large range of $t^\prime$. Interestingly the CLAS collaboration 
\ci{kubarowsky} has measured the $\eta/\pi^0$ cross section ratio for $-t$ 
reaching from $0.14$ to $0.77\,\gev^2$ but for values of $Q^2$ and $W$ that
are somewhat too low for allowing a comparison with our results without 
reservation. Nevertheless the preliminary CLAS data match very well our 
predictions which can be regarded as an indication of large contributions 
from the GPD $\bar{E}_T$ with the same sign for its $u$ and $d$-quark parts. 
We note in passing that our $\eta/\pi^0$ ratio of the longitudinal cross
sections at low $-t^\prime$ is in agreement with an estimate presented in 
\ci{eides99}. 
  
The transverse cross section for $\eta^\prime$ production is much smaller
than those for the $\eta$ and $\pi^0$ channels. The reason is obvious. The 
twist-3 mechanism is not enhanced by the chiral condensate for the 
flavor-singlet part. Actually it is about half as strong as for the octet 
channels. 

\begin{figure}[t]
\begin{center}
\includegraphics[width=0.45\tw]{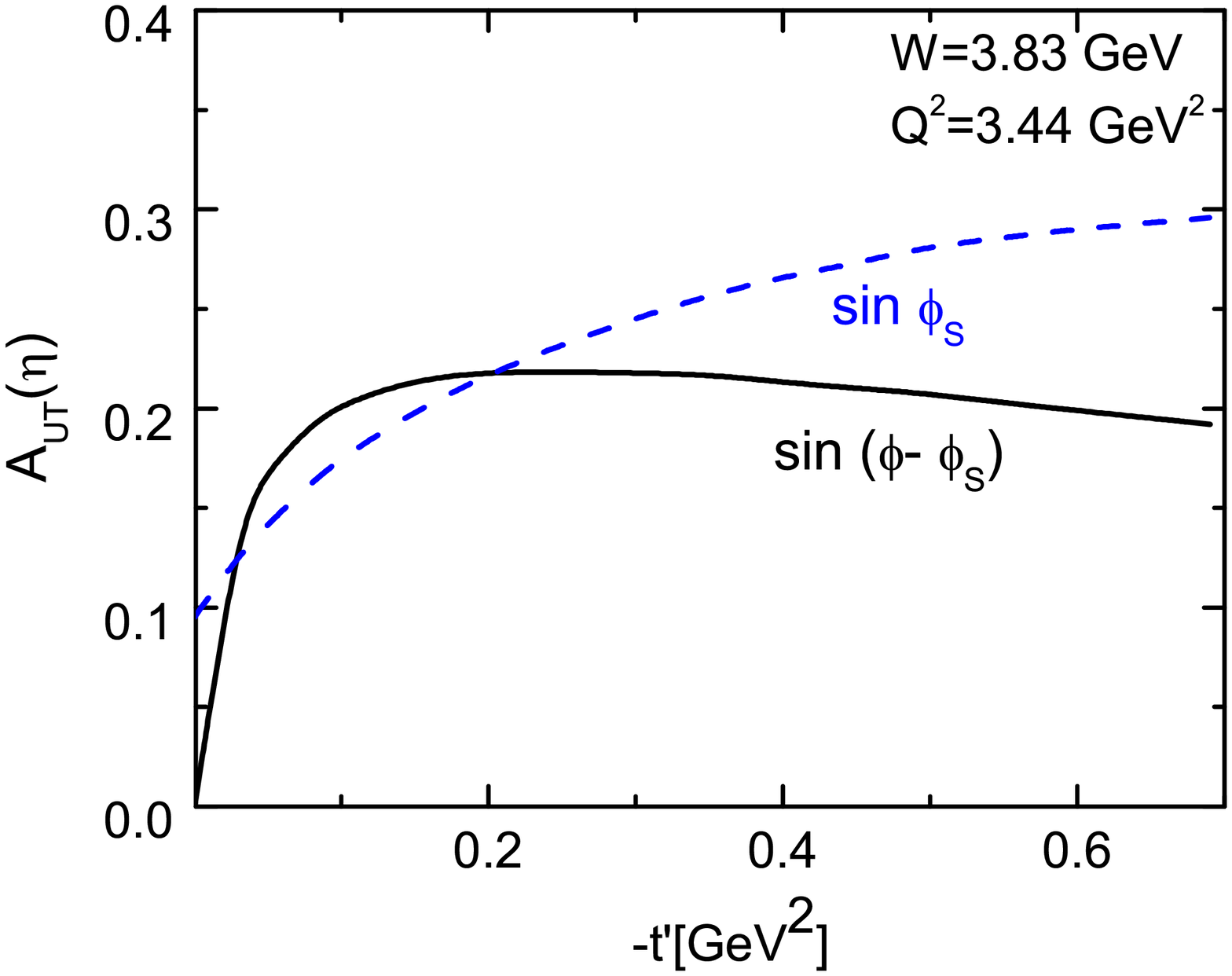} 
\includegraphics[width=0.45\tw]{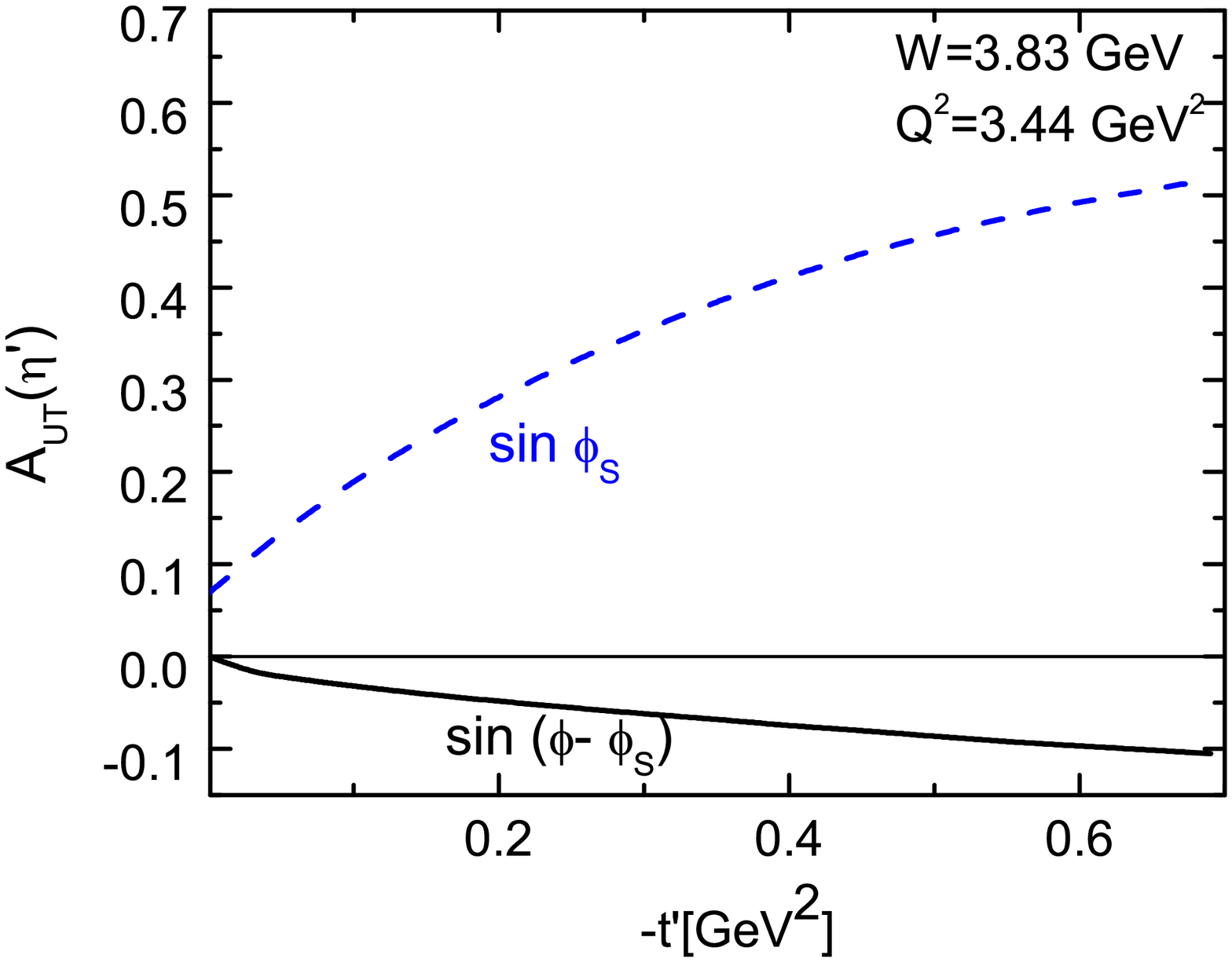}
\caption{Results for the $\sin{(\phi-\phi_s)}$ and $\sin{\phi_s}$ moments of
  the transverse target asymmetries for $\eta$ (left) and $\eta^\prime$ (right)
  electroproduction versus $t^\prime$.  (Color online)}
\label{fig:aut-eta-etap}
\end{center}
\end{figure}

In Fig.\ \ref{fig:aut-eta-etap} the $\sin{(\phi-\phi_s)}$ and the
$\sin{\phi_s}$ moments of the $\eta$ and $\eta^\prime$ cross sections measured
with a transversely polarized target are shown. The trends of these
asymmetries for $\eta$ production bear similarities to the corresponding
$\pi^0$ ones while for the $\eta^\prime$ the suppression of the twist-3 effect 
reflects itself in a different behavior of the asymmetries. 

\section{The kaon-hyperon channels}
\label{sec:kaon}
The analysis of kaon electroproduction is similar to $\pi^+$ production; 
the same expressions hold for the convolutions ($e_d=e_s$). 
To describe the $K$-meson we again use a Gaussian \wf{}  with a 
transverse-size parameter $a_K=a_\pi$ and associated distribution amplitudes,
the flat one at twist-3 level \ci{ball06} and  
\be
\Phi_K(\tau,\mu)=\Phi_{AS}(\tau)\Big[ 1 + a_{K1}(\mu)C^{3/2}_1(2\tau-1) +
                  a_{K2}(\mu)C^{3/2}_2(2\tau-1) + \ldots\Big]
\label{eq:KDA}
\ee
for twist 2 (the momentum fraction $\tau$ refers to the $u$ quark).
The important difference to the case of pions is that this distribution
amplitude is not symmetric in the momentum fractions $\tau$ and $1-\tau$ in
general. A restriction to the asymptotic distribution amplitude is
inappropriate for kaons since it would lead to the relation 
\be
{\cal H}^K \= f_K/f_\pi {\cal H}^\pi
\ee
among the twist-2 subprocess amplitudes. With $f_K=1.21 f_\pi$ the kaon 
amplitudes would then be larger than the pion ones. This is against
experience; kaon channels are typically suppressed by about $10\%$ against 
pion channels. This can be seen, for instance, in the electromagnetic form 
factors \ci{cleo-FF}, in two-photon annihilations \ci{belle} or $\chi_{cJ}$
decays into pairs of kaons or pions \ci{pdg}. For the lowest two Gegenbauer 
coefficients we use the values
\be
a_{K1}(1\,\gev)\=-0.1\,, \qquad a_{K2}(1\,\gev)\=-0.1\,,
\label{eq:gegenbauer}
\ee 
which are in agreement with the $\chi_{cJ}\to K\bar{K}$ decay width ($J=0, 2$)
\ci{BKS2}. QCD sum rules \ci{ball06}, on the other hand, provide 
$a_{K1}\=-0.06\pm 0.03$ and $a_{K2}\=0.30\pm 0.15$. The negative values of
$a_{K1}$ imply that, as expected, the antistrange quark in the kaon carries a 
larger momentum fraction on the average than the non-strange one 
($\langle \tau\rangle=1/2+3/10 a_{K1}$; $\langle 1-\tau\rangle=1/2-3/10 a_{K1}$). 
We take the first two Gegenbauer terms with the coefficients \req{eq:gegenbauer}, 
into account in order to avoid an overestimate of kaon production. Their
suppression within the modified perturbative approach (cf.\ the discussion in 
Sect.\ \ref{sec:pion}) is mild. For the twist-3 distribution amplitude we
ignore such an effect. 

For kaon production the proton-hyperon transition GPDs enter which, with the
help of SU(3) flavor symmetry, can be related to the proton GPDs by 
\ci{frankfurt99} 
\ba
F_{i p\to\Lambda\,}&\simeq& -\frac1{\sqrt{6}}\big[2F_i^u - F_i^d -
F_i^s\big]\,,\nn\\
F_{i p\to \Sigma^0} &\simeq& -\frac1{\sqrt{2}}\big[F_i^d - F_i^s\big]\,, \nn\\
F_{i p\to\Sigma^+} &\simeq& -F_i^d + F_i^s\,,
\label{eq:ptoY}
\ea
As for the other channels we assume a flavor-symmetric sea for all GPDs and 
$F_i^s-F_i^{\bar{s}}\simeq 0$. In contrast to the case of the $\pi^+$ where the
transition GPDs $F_{i p\to n}$ are related to the proton ones by isospin 
invariance, one has to be aware of possible large flavor-symmetry breaking 
effects in \req{eq:ptoY}. 

Finally, we have to discuss the kaon pole. It is treated in analogy to the
pion pole (see left hand graph in Fig.\ \ref{fig:graphs}). The contribution of
the kaon pole to $\widetilde{E}_{p\to Y}$ is given in \req{eq:Etilde-pole}
and that to the $p\to Y$ form factor reads
\be
F_{KY}^{\rm pole}(t) \= \frac{(m+M)f_K g_{KpY}}{m_K^2-t}\,F_{KpY}(t)\,,
\label{eq:kaon-pole}
\ee
where the form factor $F_{KpY}$ parameterizes a residual $t$ dependence like
the one which occurs for the pion. For this form factor which has not been
considered in previous work (see for instance \ci{bel-rad}), one may take the
same expression as for the case of the pion \req{eq:Fpipn} for a first estimate.
The GPD $\widetilde{E}$ satisfies the sum rule \ci{bel-rad}
\be
\int_{-1}^1 dx \widetilde{E}_{p\to Y}(x,\xi,t) \= F_{KY}(t) +
\frac1{\xi} g_2^{KY}(t)\,.
\label{eq:E-sum-rule}
\ee 
Due to violation of flavor symmetry by the strange quark mass the extra form
factor $g_2^{KY}$ occurs in the sum rule. This of no relevance here since
the sum rule \req{eq:E-sum-rule} will not be exploited. It is important to 
realize that in \req{eq:Etilde-pole} and \req{eq:E-sum-rule} the
transition GPD actually appears, no assumption on flavor symmetry is required.
For the kaon-baryon coupling constants appearing in \req{eq:kaon-pole} we use 
values obtained from fixed-t dispersion relations which are however in good 
agreement with SU(3) predictions \ci{compilation}
\be
g_{K^+p\Lambda}\simeq-13.3\,, \qquad 
                  g_{K^+p\Sigma^0}\=-g_{K^0p\Sigma^+}/\sqrt{2}\simeq-3.5\,.
\ee
The relation between the coupling constants for $K^+p\Sigma^0$ and 
$K^0 p \Sigma^+$ is a consequence of isospin invariance. The residue of the
kaon pole \req{eq:residue} is fully fixed by noting that we take for the
electromagnetic form factor of the positively charged kaon that of the pion
but multiplied by a factor of 0.9 in order to take care of flavor symmetry
breaking
\be
F_{K^+}(Q^2) \= \frac{0.9}{1+Q^2/0.462\,\gev^2}\,.
\ee
This is at least in agreement with the CLEO measurement \ci{cleo-FF} of 
the time-like pion and kaon electromagnetic form factors. 

The electromagnetic form factor of the $K^0$ is not necessarily zero. In the 
perturbative frame work for instance it is related to the antisymmetric 
Gegenbauer coefficients in \req{eq:KDA}. Since nothing is known about this 
form factor we assume its ratio to the $K^+$ form factor to be given by the 
perturbative ratio 
\be
F_{K^0}/F_{K^+}\approx \frac43\,\frac{a_{K1} (1+a_{K2})}
               {1+a_{K1}^{2}+a_{K2}^{2} + 2 a_{K2}}\,.
\ee
Evaluating this ratio from the Gegenbauer coefficient quoted in
\req{eq:gegenbauer}, one finds a value of -0.15.

\begin{figure}[t]
\begin{center}
\includegraphics[width=0.32\tw]{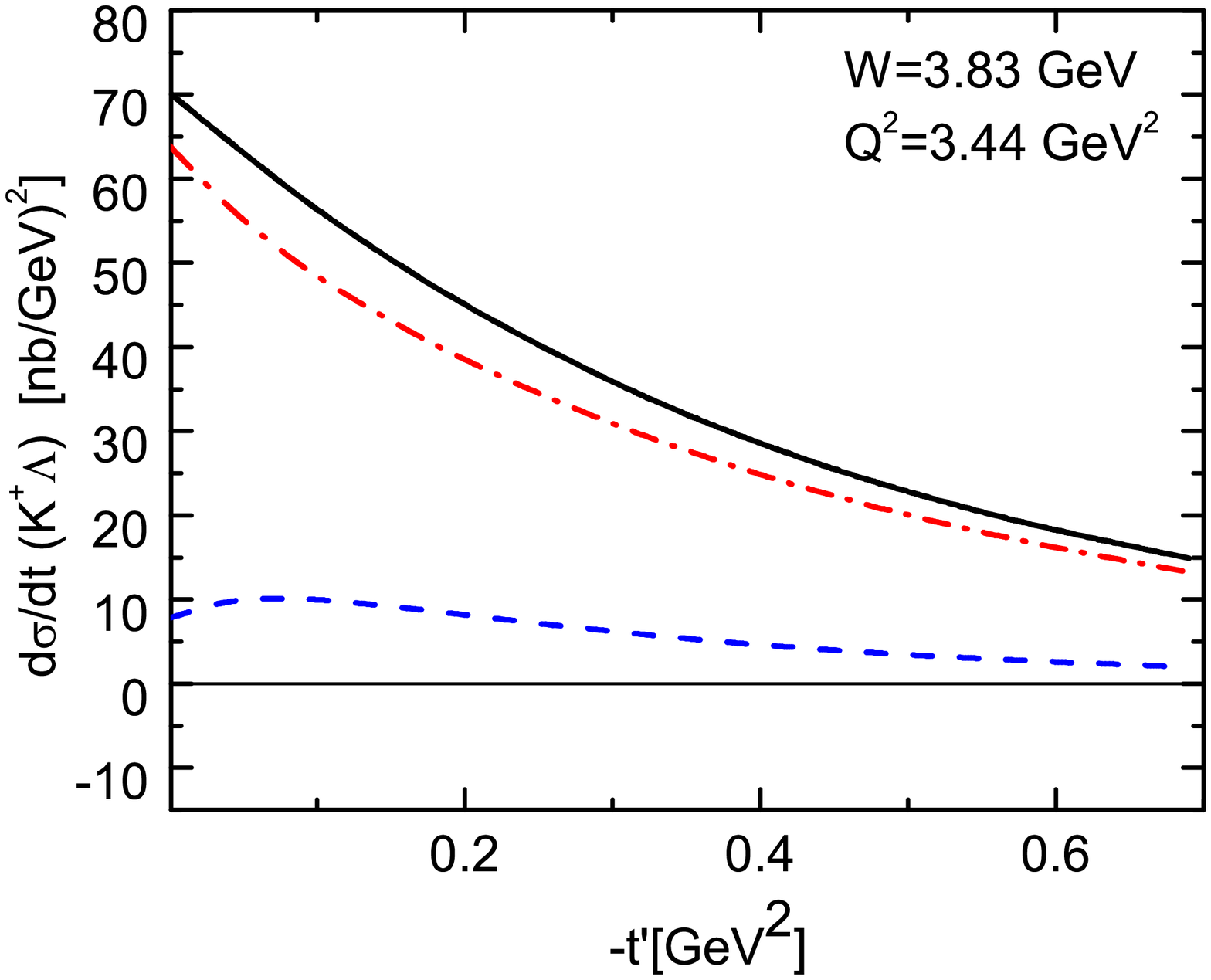} 
\includegraphics[width=0.32\tw]{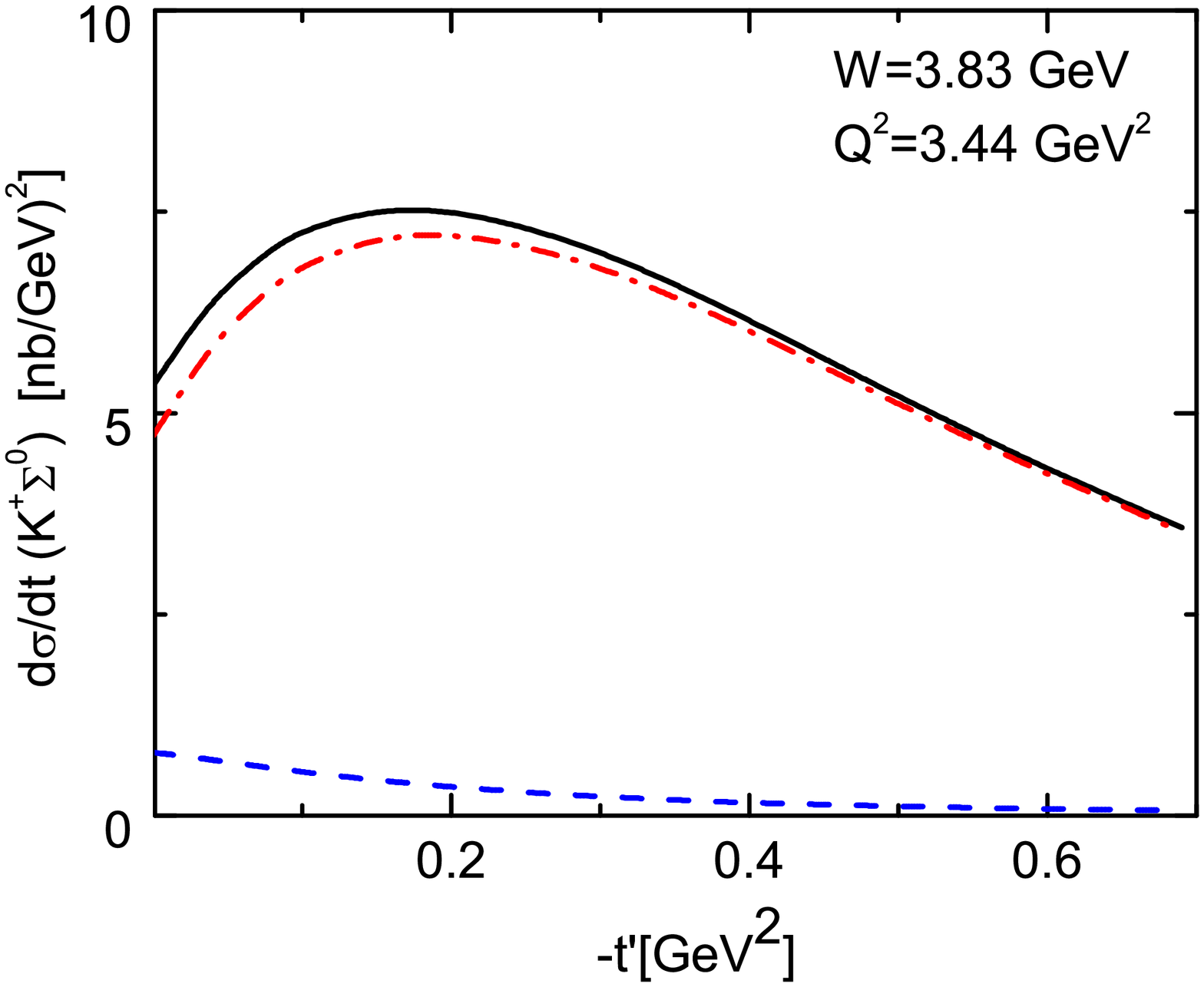}
\includegraphics[width=0.32\tw]{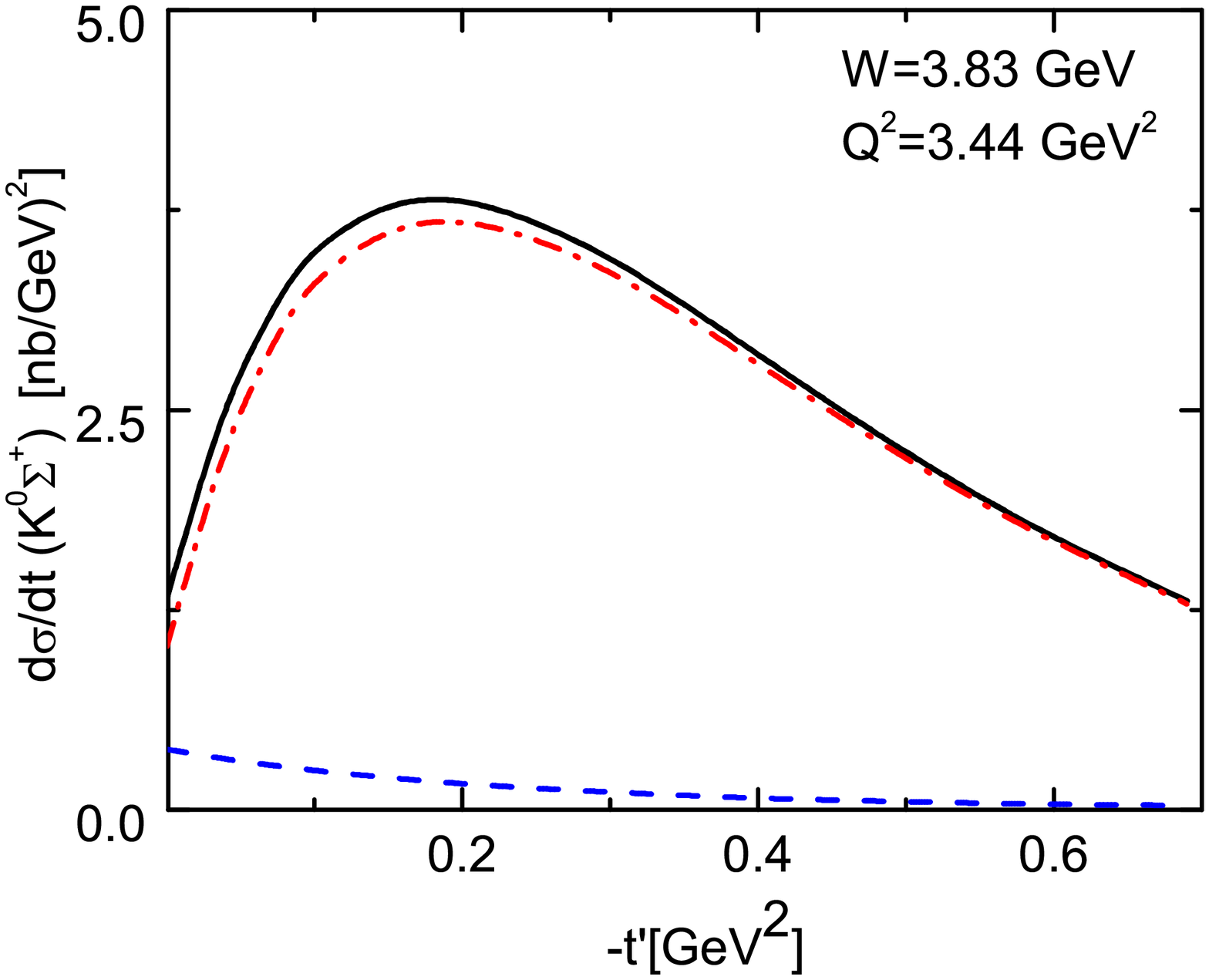}
\caption{Results for the cross sectios of the $K^+\Lambda$ (left), 
$K^+\Sigma^0$ (center) and $K^0\Sigma^+$ (right) final states. For 
notations refer to Fig.\ \ref{fig:sigma-pi+-pi0}. (Color online)}  
\label{fig:sigma-kaon}
\vspace*{0.03\tw}
\includegraphics[width=0.32\tw]{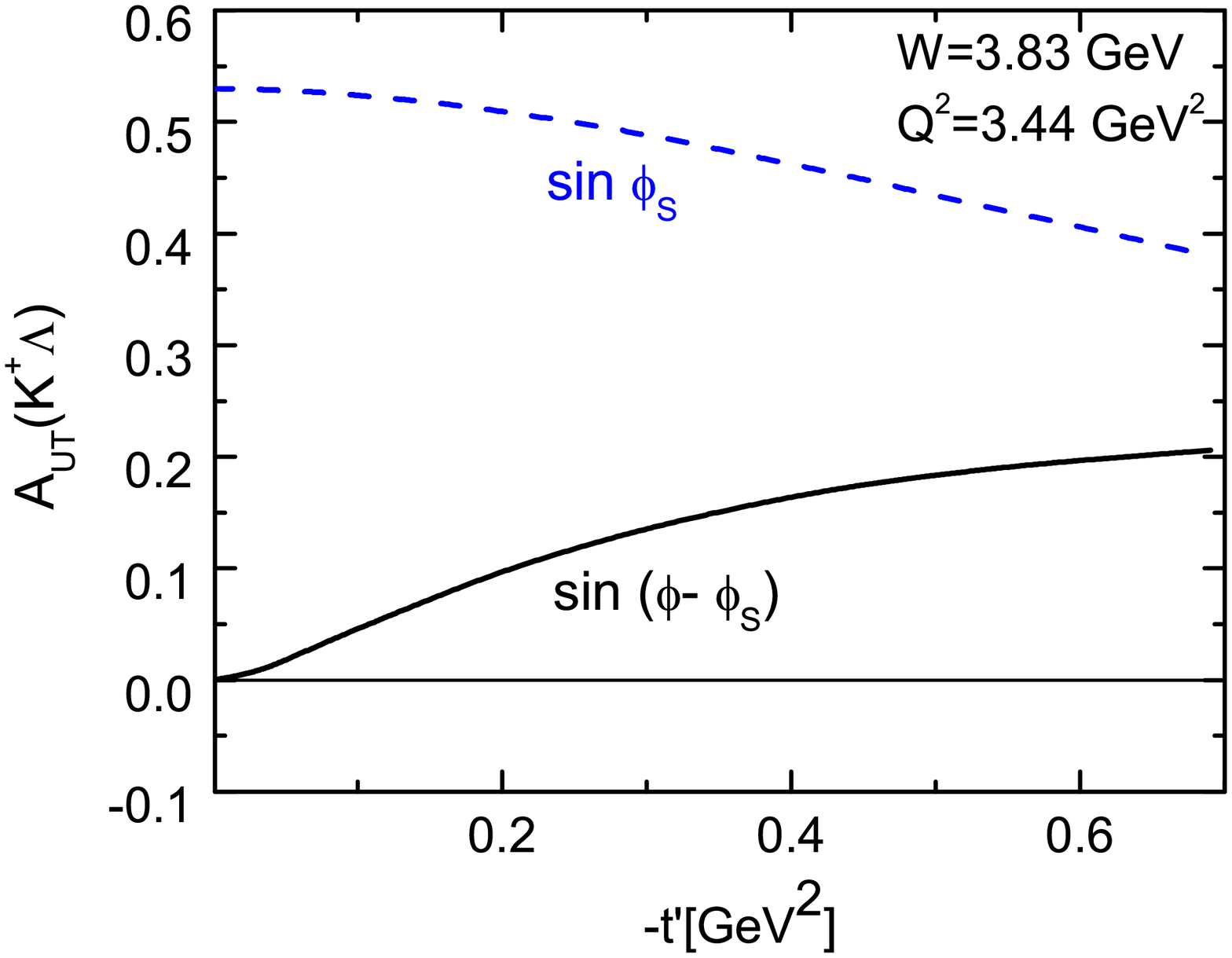} 
\includegraphics[width=0.32\tw]{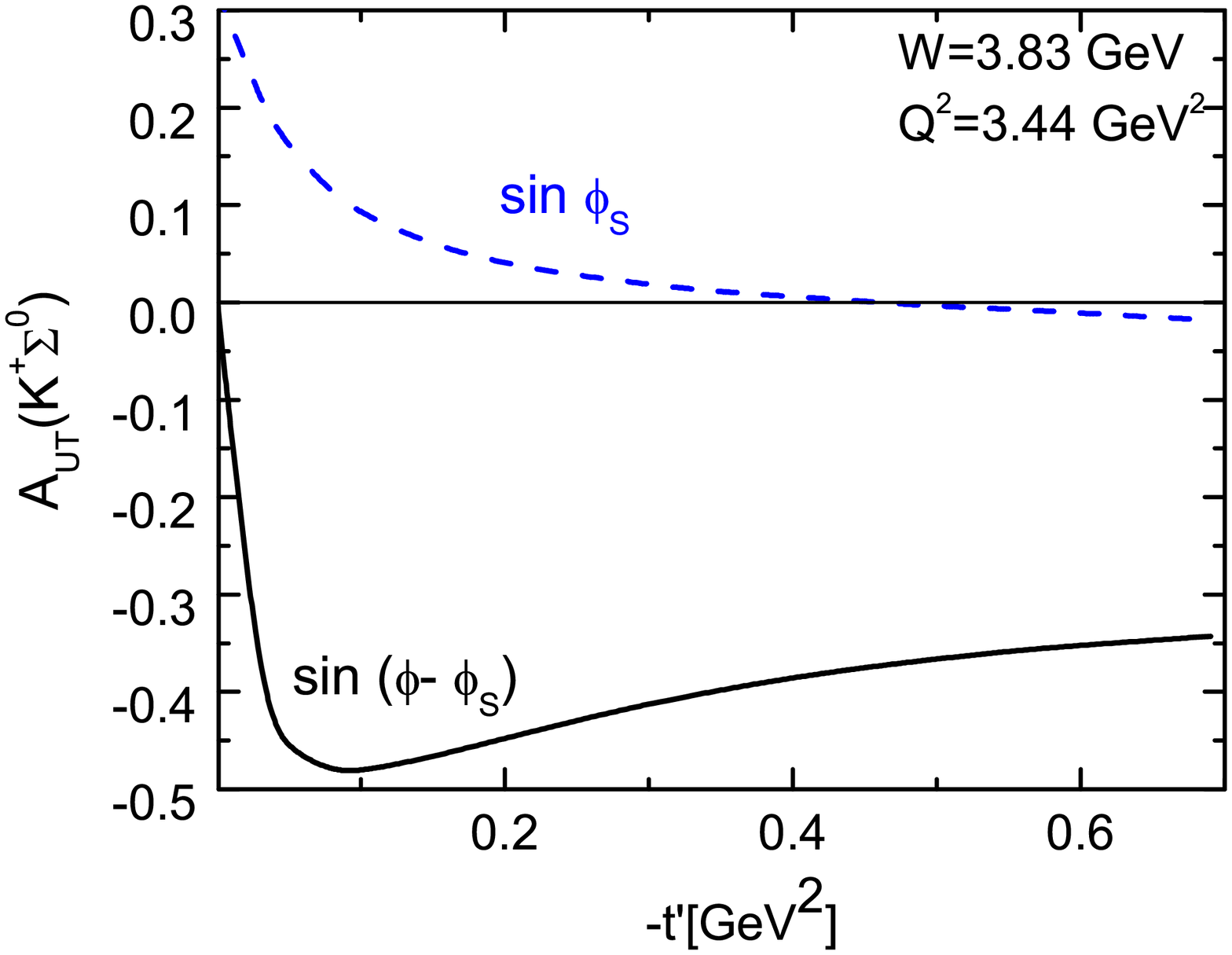}
\includegraphics[width=0.32\tw]{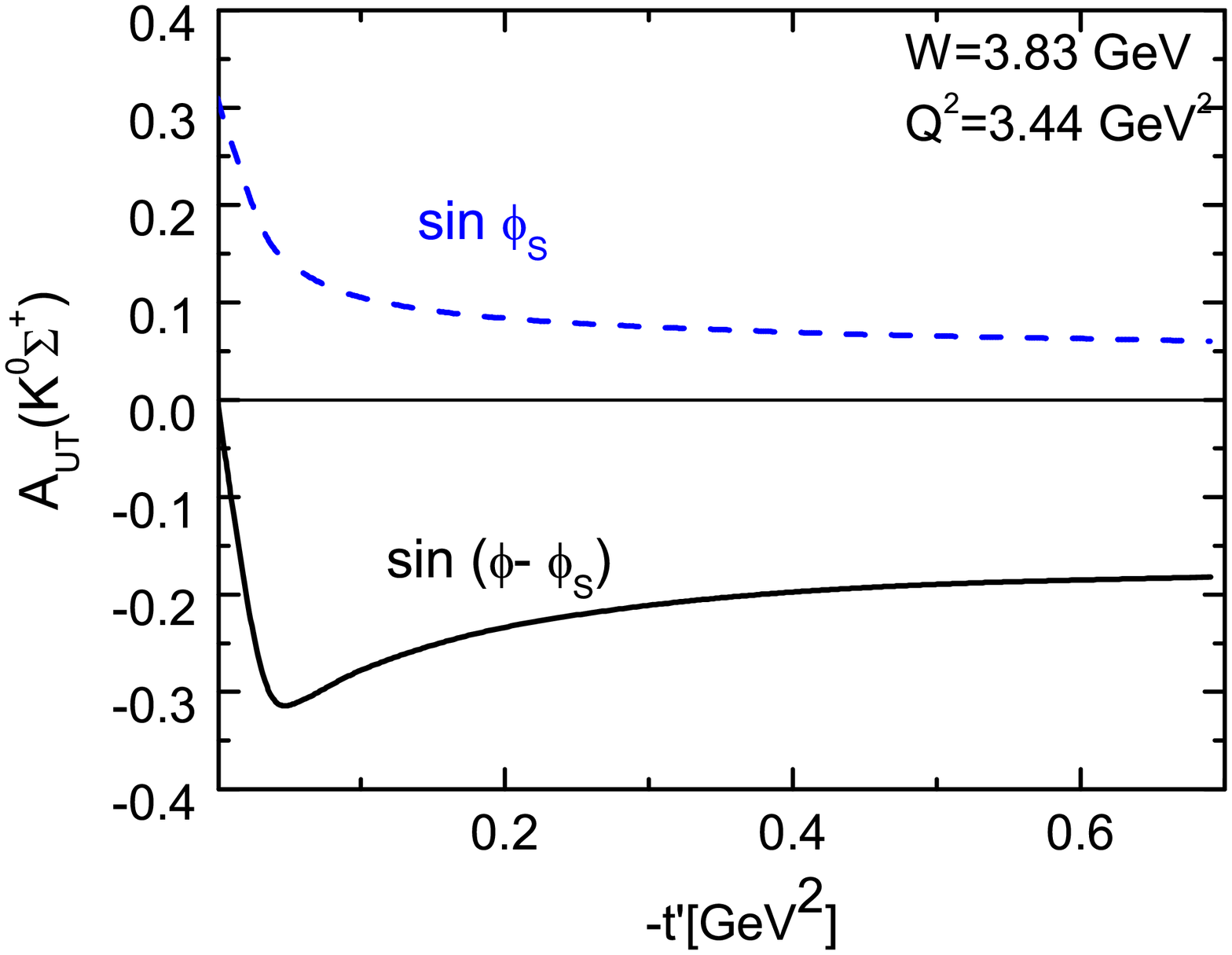}
\caption{Results for the $\sin{(\phi-\phi_s)}$ and $\sin{\phi_s}$ moments
of the transverse target asymmetries for electroproduction of the $K^+\Lambda$ 
(left), $K^+\Sigma^0$ (center) and $K^0\Sigma^+$ (right) final states
versus $t^\prime$. (Color online)}
\label{fig:aut-kaon}
\end{center}
\end{figure}

Our results for cross sections and target asymmetries are shown in Figs.\
\ref{fig:sigma-kaon} and \ref{fig:aut-kaon}. The kaon cross sections
are dominated by contributions from transversely polarized photons too. As
opposed to $\pi^+$ production the meson-pole term plays a minor role 
here because of the large kaon mass appearing in the pole denominator.
It is however strong enough to produce a little maximum in the longitudinal 
cross section for $K^+\Lambda$ at $t^\prime\simeq -0.1\,\gev^2$ by means of 
the factor $-t/(t-m_K^2)^2$ in the pure pole contribution ($m_K^2>-t_0$). For 
the two other kaon channels the pole contribution is hardly visible. The 
cross section for the $K^+\Lambda$ channel is substantially larger than those 
for the other kaon channels. Besides different pole contributions, these 
two processes differ from each other by the factor $1/\sqrt{2}$ in the GPD 
combinations \req{eq:ptoY} in favor of $\gamma^*p\to K^0\Sigma^+$. This factor 
is overcompensated by the $u$-quark charge in the hard amplitude of the
process $\gamma^*p\to K^+\Sigma^0$ instead of $e_d$ in the other process. As 
a result the $K^+\Sigma^0$ cross section is about twice as large as the 
$K^0\Sigma^+$ one. On the other hand, the transverse target asymmetries 
behave very similar for both the processes, see Fig.\ \ref{fig:aut-kaon}. 
The asymmetries for the $K^+\Lambda$ channel are similar to those for $\pi^+$
production.

\begin{figure}[t]
\begin{center}
\includegraphics[width=0.45\tw]{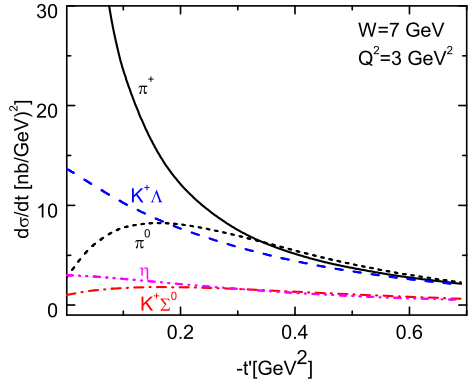}  
\caption{The cross sections for various pseudoscalar meson channels at
  $W=7\,\gev$ and $Q^2=3\,\gev^2.$ (Color online)}  
\label{fig:ds-all}
\end{center}
\end{figure}

Finally, in Fig.\ \ref{fig:ds-all}, we compare the various cross sections at a
larger value of $W$, characteristic of the COMPASS experiment.

\section{The longitudinal beam and target\\ asymmetries}
\label{sec:supp}
\begin{figure}[t]
\begin{center}
\includegraphics[width=0.45\tw]{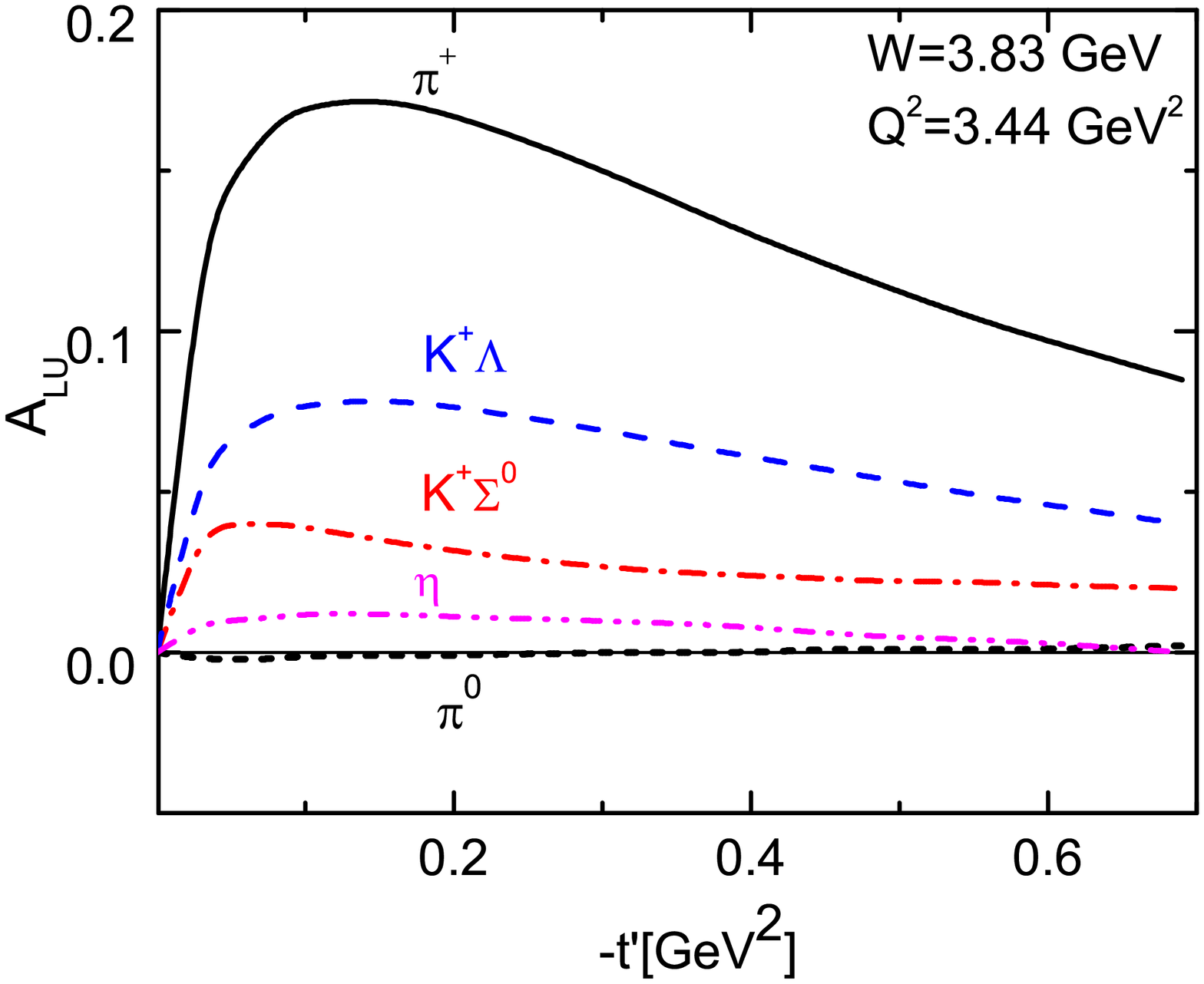} 
\includegraphics[width=0.45\tw]{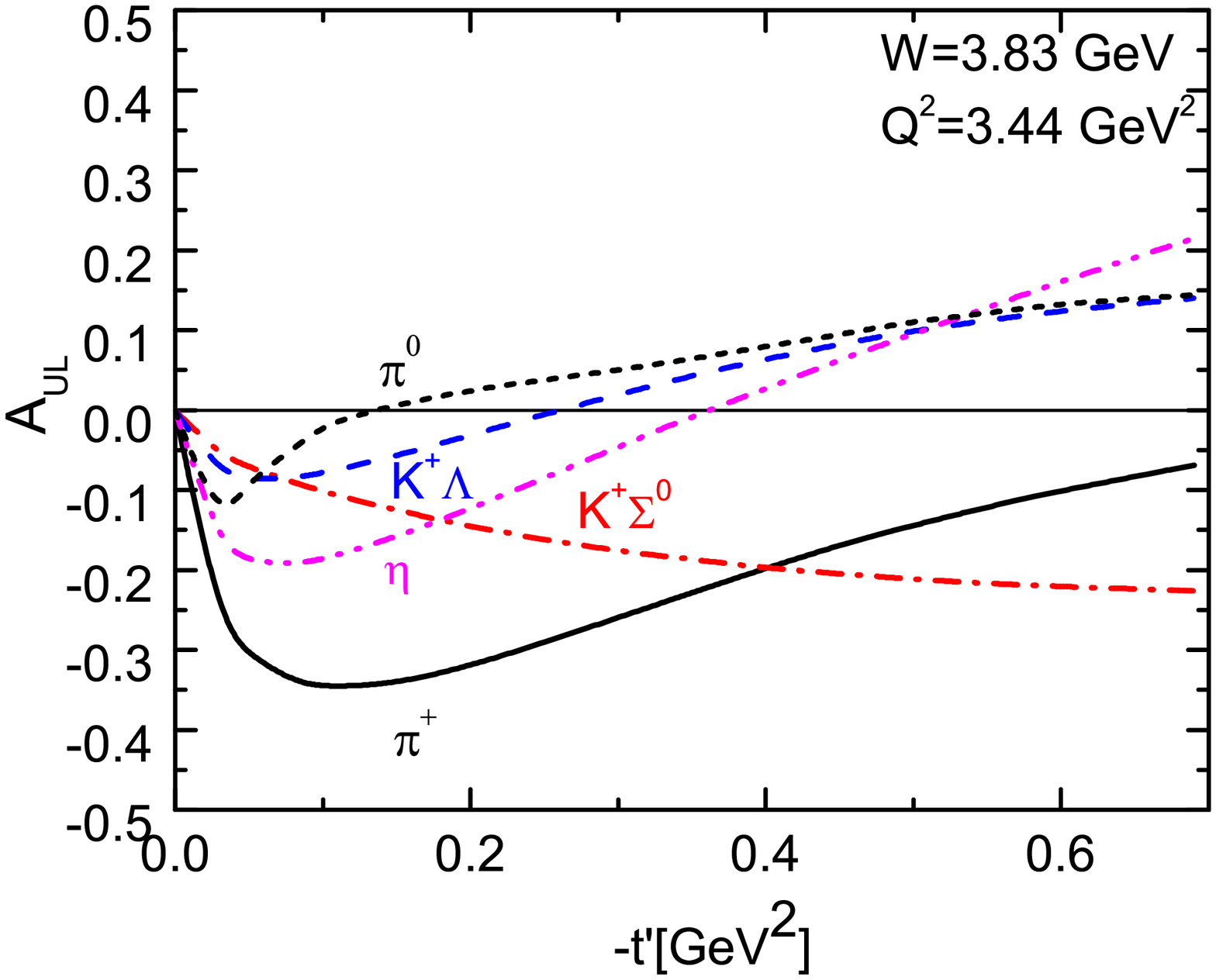} 
\caption{The beam spin asymmetry (left) and the asymmetry for a longitudinally
  polarized target (right) for various pseudoscalar-meson channels 
  versus $t^\prime$. (Color online)}  
\label{fig:alu}
\end{center}
\end{figure}

In Fig.\ \ref{fig:alu} we display the beam-spin asymmetry for the
pseudoscalar-meson channels at the standard kinematical setting. In terms of
helicity amplitudes this observable is expressed by
\ba
A_{LU}\sigma_0 &=& \sqrt{\varepsilon (1-\varepsilon)} {\rm Im}
\Big[\big({\cal M}^*_{0+,++} - {\cal M}^*_{0+,-+}\big) {\cal M}_{0+,0+}\nn\\
&+& \big({\cal M}^*_{0-,++} - {\cal M}^*_{0-,-+}\big) {\cal M}_{0-,0+}\Big]\,,
\label{eq:alu}
\ea
In our handbag approach the first term in \req{eq:alu} cancels (see
\req{eq:ebar-ampl}) and the second one simplifies to 
\be 
A_{LU}\sigma_0 \simeq \sqrt{\varepsilon (1-\varepsilon)} {\rm Im}
 \Big[{\cal M}^*_{0-,++}\, {\cal M}_{0-,0+}\Big]\,.
\ee
For reactions being dominated by $\bar{E}_T$ as for example $\pi^0$ 
production, $A_{LU}$ is very small. Others like $\pi^+$ production 
for which $\bar{E}_T$ is less prominent, are large. This can be 
clearly seen in Fig.\ \ref{fig:alu}. The CLAS collaboration \ci{harut}
has measured $A_{LU}$ for $\pi^0$ and $\pi^+$ production at $W\simeq 2.4\,\gev$
and $Q^2\simeq 2.5\,\gev^2$. While the CLAS result for $\pi^+$ production 
is in reasonable agreement with our result, is that for $\pi^0$ production 
larger than our one.

Another interesting observable is the $\sin{\phi}$-moment of the
electroproduction cross section measured with a longitudinally polarized 
target. To the extend that the angle of the rotation in the lepton plane 
from the direction of the incoming lepton to the virtual photon is small
the longitudinal target asymmetry reads
\ba
A_{UL}\sigma_0&\simeq& -\sqrt{\varepsilon(1+\varepsilon)}{\rm Im}  
\Big[\big({\cal M}^*_{0+,++} + {\cal M}^*_{0+,-+}\big)\, {\cal M}_{0+,0+}\nn\\
&&\quad +\, \big({\cal M}^*_{0-,++} + {\cal M}^*_{0-,-+}\big)\, 
{\cal M}_{0-,0+}\Big]\,.
\label{eq:aul}
\ea
It simplifies in our handbag approach to
\ba
A_{UL}\sigma_0&\simeq& -\sqrt{\varepsilon(1+\varepsilon)}{\rm Im}
\Big[2{\cal M}^*_{0+,++}\,{\cal M}_{0+,0+}\nn\\
&&\quad + \,{\cal M}^*_{0-,++}\, {\cal M}_{0-,0+}\Big]\,.
\ea
Without the GPD $\bar{E}_T$ the ratio of the two asymmetries would be given
by $A_{UL}/A_{LU}=-\sqrt{(1+\varepsilon)/(1-\varepsilon)}$. Deviations
from this ratio signal a non-zero contributions from $\bar{E}_T$, see Fig.\ 
\ref{fig:alu}.
\section{Summary}
\label{sec:summary}
We have investigated electroproduction of pseudoscalar mesons within the 
handbag approach restricting ourselves to the range of small skewness. 
Forced by the behavior of target asymmetries for $\pi^+$ production
measured by the HERMES collaboration \ci{Hristova} we have allowed for a
twist-3 effect consisting of a twist-3 meson distribution amplitude in
combination with the leading-twist transversity GPDs. This twist-3 effect is 
strongly enhanced by the chiral condensate \req{eq:chiral}, \req{eq:chiral-eta}.
It therefore seems to be justified to include this effect in the analysis  
even though we don't perform a systematic analysis at twist-3 accuracy.
We stress that for electroproduction of vector mesons the twist-3 effect
is expected to be much smaller since $\mu_P$ is to be replaced by the vector
meson's mass.

The GPDs are constructed from double distributions which are parameterized as
zero-skewness GPDs multiplied by a Regge-like $t$ dependence and a weight
function from which the skewness dependence of the GPD is generated. They
are constrained by form factors and PDFs. Their isovector combination are
fitted to the HERMES data on $\pi^+$ electroproduction. Moments of the GPDs 
are compared to moments obtained from lattice QCD. In general reasonable 
agreement is to be observed. A number of positivity bounds for the GPDs are 
checked as well by us. An exceptional case is $\bar{E}_T$. Its
parameterization is fully constrained by lattice-QCD results \ci{goeckeler} 
since this is the only available information about it at present. Therefore, 
$\bar{E}_T$ suffers from the usual uncertainties of lattice-QCD results 
arising from the still low statistics and from the compulsion of working with 
heavy quarks. It turns out that the GPD $\bar{E}_T$ for $u$ 
and $d$ quarks cancel to a high degree in $\pi^+$ electroproduction while 
it plays a prominent role in most of the other pseudoscalar-meson channels. In
fact, with the exception of $\pi^+$ and $\eta^\prime$ they are all dominated
by contributions from the transversity GPDs, i.e.\ by transversely polarized
photons, in the range of $Q^2$ accessible in present-day experiments. Of
course, for $Q^2\to\infty$ the longitudinal cross sections will dominate in
all reactions. These findings are to be contrasted with the usual assumption 
made in the handbag approach \ci{frankfurt99}-\ci{bechler} that contributions 
from longitudinally polarized photons already dominate for $Q^2$ of the order 
of a few GeV. 
 
A precise calculation of all details of electroproduction of pseudoscalar
mesons is beyond feasibility at present. There are many uncertainties 
like the parameterization of the transversity GPDs or the exact treatment 
of the twist-3 contribution (e.g.\ the neglect of possible three-particle 
configuration of the meson state). Also higher-order perturbative 
corrections other than those included in the Sudakov factor and, implicitly, 
in the experimental electromagnetic form factors of the mesons, are ignored. 
According to \ci{kugler} the NLO corrections are rather large for the cross 
sections for $Q^2\lsim 10\,\gev$. However, the results presented in
\ci{kugler} refer to the leading-twist contribution to $\pi^+$ production, 
i.e.\ to about $10\%$ of the cross section since only the perturbative 
contribution to the pion form factor is taken into account. Thus, it is not
clear how strong the NLO corrections are in our approach. Further
uncertainties occur for kaon production. In contrast to the case of the $\pi^+$ 
where the $p\to n$ transition GPDs are related to the diagonal proton ones by 
isospin symmetry, the proton - hyperon transition GPDs are connected to the 
proton GPDs by SU(3) flavor symmetry which is less accurate than isospin 
symmetry. The assumption of a flavor symmetric sea for all GPDs is also 
stronger for kaons than for pions. With regard to all these uncertainties 
we consider our investigation of electroproduction of pseudoscalar mesons 
as an estimate of the pertinent observables. The trends and magnitudes of 
the cross sections are likely correct but probably not the details. 
Particularly large uncertainties are to be expected for subtle observables 
as the spin asymmetries. On the other hand, the large magnitude of 
contributions from the transversity GPDs seems to us hard to avoid. Future 
data measured at the upgraded Jefferson Lab or by the COMPASS collaboration 
may probe our results. Although our approach is designed for small skewness 
(and rather large $Q^2$ and $W$) we also compared our results with the large 
skewness data measured at the present JLab accelerator. We found general 
agreement in trends and magnitudes. We regard this as a hint at strong
contributions from the transversity GPD $\bar{E}_T$ for which the $u$ and $d$
quark parts have the same sign.

{\bf Acknowledgements}  We thank Ph.\ H\"agler for drawing our attention to
\ci{goeckeler}. We are also grateful to H.\ Avagyan, M.\ Diehl, V.\ Kubarowsky, 
A.\ Metz and W.-D.\ Nowak for valuable discussions.
This work is supported  in part by the Russian Foundation for
Basic Research, Grant 09-02-01149 and by the Heisenberg-Landau
program and by the BMBF, contract number 06RY258.\\


\end{document}